\documentclass[10pt,journal,compsoc]{IEEEtran}
\usepackage[nocompress]{cite}
\usepackage{graphicx}
\usepackage{amsmath, amssymb}
\usepackage{url}
\usepackage{booktabs}                  
\usepackage{lipsum}                    
\usepackage{mwe}                       
\usepackage{graphicx}
\usepackage{multirow}
\usepackage{hyperref}
\usepackage{array} 
\usepackage[caption=false,font=footnotesize]{subfig}

\begin{document}

\title{Interactive Hypergraph Visual Analytics for Exploring Large and Complex Image Collections}

\author{%
  Floris~Gisolf, %
  Zeno~J.~M.~H.~Geradts,~\IEEEmembership{Senior Member, IEEE} %
  and Marcel~Worring,~\IEEEmembership{Senior Member, IEEE}%
  \IEEEcompsocitemizethanks{%
    \IEEEcompsocthanksitem Floris Gisolf, Zeno J.~M.~H.~Geradts, and Marcel Worring are with the University of Amsterdam, The Netherlands.
    E-mail: \{f.gisolf, z.j.m.h.geradts, m.worring\}@uva.nl.%
    \IEEEcompsocthanksitem Floris Gisolf is also with the Dutch Safety Board, The Hague, The Netherlands.%
    \IEEEcompsocthanksitem Zeno J.~M.~H.~Geradts is also with the Netherlands Forensic Institute, The Hague, The Netherlands.%
    \IEEEcompsocthanksitem Code available at \url{https://github.com/Fenre1/HypergalleryQ}.%
  }%
}

\IEEEtitleabstractindextext{%
\begin{abstract}
Analyzing large complex image collections in domains like forensics, accident investigation, or social media analysis involves interpreting intricate, overlapping relationships among images. Traditional clustering and classification methods fail to adequately represent these complex relationships, particularly when labeled data or suitable pre-trained models are unavailable. Hypergraphs effectively capture overlapping relationships, but to translate their complexity into information and insights for domain expert users visualization is essential. 
We propose an interactive visual analytics approach specifically designed for the construction, exploration, and analysis of hypergraphs on large-scale complex image collections. Our core contributions include: (1) a scalable pipeline for constructing hypergraphs directly from raw image data, including a similarity measure to evaluate constructed hypergraphs against a ground truth, (2) interactive visualization techniques that integrate spatial hypergraph representations, interactive grids, and matrix visualizations, enabling users to dynamically explore and interpret relationships without becoming overwhelmed and disoriented, and (3) practical insights on how domain experts can effectively use the application, based on evaluation with real-life image collections. Our results demonstrate that our visual analytics approach facilitates iterative exploration, enabling domain experts to efficiently derive insights from image collections containing tens of thousands of images.
\end{abstract}
}
\maketitle
\IEEEdisplaynontitleabstractindextext  

\begin{center}
\textit{This paper has been submitted to IEEE Transactions on Visualization and Computer Graphics for review.}
\end{center}

\section{Introduction}
\IEEEPARstart{I}{n} domains such as forensics and accident investigation, experts have to make well informed and high impact decisions with access to unannotated complex image collections only. A complex image collection (CIC) typically contains thousands to as many as 100,000 images or more, each depicting multiple objects, captured at specific times and locations, and various types of relations among them. Throughout this paper, CIC specifically refers to image collections that are unannotated.

Two characteristics make CICs particularly challenging to analyze. First, they contain many overlapping and non-exclusive relationships: a single image may contain multiple objects, and images can be grouped based on similarities in their content, or spatial, temporal, and environmental context. A ground truth annotation for such an image collection would therefore necessarily be multi-label. Second, their unique and specialized content creates a semantic gap, causing standard pre-trained models to perform poorly. Bridging this gap with common techniques like fine-tuning is often impossible, as it requires annotated data. In contrast, a visual analytics approach supports expert-driven exploration and interpretation of CICs without requiring large-scale annotation. A flexible representational framework is needed that can accommodate both the domain-specific content and the complex overlapping relationships among images. This framework can serve either as a standalone decision-support tool for experts or for annotating the collection to enable subsequent fine-tuning. 

Complex, overlapping and non-exclusive relationships are a challenge that is also present in social network analysis, where individuals frequently belong simultaneously to multiple groups or communities; hypergraphs have proven particularly effective in modeling these relationships\cite{Battiston2020HigherOrderRepresentations}. Hypergraphs\cite{Berge1984HypergraphsCombinatoricsFinite}, which allow elements to participate in multiple clusters simultaneously, provide the necessary flexibility to enable a visual analytics solution for a more accurate representation of the underlying relationships within the data. 

To develop our visual analytics solution, we need to understand how domain experts analyze large CICs. Typically, experts start with unstructured collections and gradually organize them into meaningful structures through iterative exploration and targeted searches for relevant items, enabling essential insights \cite{Zahalka2014TowardsInteractive,Gisolf2021SearchExploreStrategies}. In this process, exploration is undertaken when the expert is faced with an unfamiliar collection and seeks to uncover the underlying structure through iterative dynamic analysis. In contrast, search is used when the expert has a clear target in mind and requires fast, precise retrieval of relevant items. Many multimedia analytics tasks, especially when analyzing CICs, require both approaches: initial exploration to understand the full scope of the data, followed by targeted searches for specific insights. Consequently, analyzing image collections involves an iterative alternation between exploration and search, and requires dynamic data exploration, filtering, querying, annotation, and hypothesis testing.

Supporting this expert-driven workflow with a hypergraph-based model presents several practical and technical challenges. The system must construct a meaningful hypergraph from raw data and also visualize it at scale. This visualization must be designed to avoid the visual clutter and cognitive overload that can easily overwhelm a user. Furthermore, it must be performant on the typical hardware available to experts, as many organizations lack access to large computer clusters or cloud resources due to cost or data confidentiality. Addressing these challenges requires a complete pipeline for hypergraph construction, scalable and digestible visualization, and quality evaluation. Our approach provides this through the following contributions:

\begin{itemize}
\item A hypergraph construction and evaluation framework: We propose a pipeline for deriving hypergraphs directly from raw image data, and outline a hypergraph similarity measure to evaluate the quality of the hypergraph;
\item Scalable hypergraph visualization: We introduce hypergraph visualization designed to scale to image collections containing up to a forty thousand images; 
\item Interactive and digestible approach: By integrating interactivity and layering information, our method enables users to gradually explore and comprehend large hypergraphs without feeling overwhelmed and disoriented, facilitating efficient and meaningful insight discovery;
\item Practical insights on how domain experts can effectively use the application, based on a structured evaluation with real-world investigative image collections.
\end{itemize}

\section{Related work}
\label{sec:related_works}
The analysis, representation, and visualization of complex large-scale datasets has garnered significant attention across various domains, with methods often tailored to the specific nature of the data. However, CICs pose unique challenges due to their unstructured nature and overlapping relationships. This section reviews existing approaches to addressing these challenges, including multi-label classification (MLC) and clustering, hypergraph construction, visualization, and evaluation.

\subsection{Measures for hypergraph similarity}

The development of any hypergraph construction method, particularly from unstructured data, cannot be meaningfully pursued without a way to evaluate its output. Robust similarity measures are essential for assessing the quality of a generated hypergraph or comparing different construction techniques. However, this remains a significant challenge: scalable and universally accepted evaluation metrics are lacking. Ground truth comparisons are especially valuable in this context, but few measures are designed specifically to compare a generated hypergraph to a ground truth counterpart.

Traditional graph similarity measures, such as Graph Edit Distance and Maximum Common Edge Subgraph, are NP-hard problems, making them computationally expensive, particularly for large graphs. To address this, deep learning approaches have been developed to approximate these methods and reduce computational costs~\cite{Yang2024DeepLearningApproaches}.
  
However, hypergraphs generalize these relationships by allowing edges (hyperedges) to connect multiple nodes, often in overlapping ways, and such higher-order structures are not easily captured by the traditional graph metrics.

In clustering evaluation, measures such as Normalized Mutual Information (NMI) and Adjusted Rand Index (ARI)\cite{Warrens2022UnderstandingAdjustedRand} are used to assess similarity between clusters. These measures are effective for datasets where each element belongs to a single, exclusive cluster. However, hypergraphs represent overlapping clusters and have multi-node hyperedges, which these traditional measures cannot fully capture.

One notable example is the Hypergraph Similarity Measure~\cite{Surana2023HypergraphSimilarityMeasures}, which uses tensor-based representations and algebraic methods to quantify structural similarity between hypergraphs. Their proposed methods suffer from a limitation: they only work when the two hypergraphs have the same maximum hyperedge size. Furthermore, their direct measures suffer from scalability issues, becoming computationally infeasible beyond approximately 100 nodes due to memory constraints. This restricts its use for large image collections, highlighting the need for more scalable and efficient measures suitable for large-scale datasets.

\subsection{Representing image collections with hypergraphs}

Complex multi-label datasets can be effectively represented using hypergraphs, as they naturally capture group relationships beyond pairs and allow elements to participate in multiple groups simultaneously. Hypergraphs, widely used in fields such as image segmentation and social network analysis\cite{Battiston2020HigherOrderRepresentations}, allow individual items to belong simultaneously to multiple groups or categories. Representing data using hypergraphs first requires a method to systematically construct the hypergraph structure from the dataset. 

Most hypergraph construction research focuses on attribute-based hypergraph generation\cite{Huang2015LearningHypergraphRegularized,Aksoy2019HighPerformanceHypergraph} or network-based hypergraph generation\cite{Fang2014TopicSensitiveInfluencer,Zu2016IdentifyingHighOrder,Franzese2019HypergraphBasedConnectivity}, where there is already a (latent) network available to generate a hypergraph~\cite{Gao2022HypergraphLearning}.

A more implicit way to construct a hypergraph would be through Multi-Label Classification (MLC)\cite{Ridnik2021AsymmetricLoss,Bogatinovski2022ComprehensiveComparative,Han2023Survey}. This approach has achieved significant success in tasks such as document categorization\cite{You2019AttentionXML,Liu2022EmergingTrends} and image classification~\cite{Lanchantin2021GeneralMultiLabel,Sovatzidi2023InterpretationML,Zhu2021ResidualAttention}. However, applying these methods to CICs is problematic. Standard pre-trained models are typically trained on generic datasets like ImageNet or MS COCO, which fail to capture the unique semantics of specialized image collections. This creates a semantic gap, leading to poor performance. While techniques like domain-specific fine-tuning or retrieval-augmented generation~\cite{Yasunaga2023RetrievalAugmentedMultimodal,Zhao2023RetrievingMultimodalInformation} can bridge this gap, they are fundamentally dependent on labeled data. Since CICs are unannotated by definition, supervised approaches like MLC are unsuitable for our use case. 

\subsection{Constructing hypergraphs from raw data}
To effectively create hypergraphs, methods that do not rely strictly on predefined categories or extensive labeled training sets become essential. Clustering provides an unsupervised alternative that can automatically uncover latent structures and capture overlapping relationships without the need for extensive labeling or retraining. However, traditional clustering methods, such as k-means, typically assign each image to a single exclusive cluster. This is overly restrictive for complex image collections that contain many overlapping relationships. 
While these methods work well for single-label datasets, they struggle to capture the overlapping relationships inherent in CICs. To address this limitation, multi-label clustering methods have been introduced, such as fuzzy c-means\cite{Askari2021FuzzyCMeansClustering} and possibilistic c-means\cite{Krishnapuram1996PossibilisticCMeans}. These methods may be a way to construct hypergraphs.


Few papers have directly addressed the construction of hypergraphs from raw data. Exceptions are HYGENE, a diffusion-based hypergraph generator\cite{gailhard2024hygenediffusionbasedhypergraphgeneration}, and HGRec++\cite{Lin2024AutomaticHypergraphGeneration}. Unfortunately, HYGENE is not suitable for our use case, as it requires a training set of hypergraphs with structural properties (such as number of nodes and hyperedges, the distributions of node degrees and hyperedge sizes, and spectral features of the hypergraph's Laplacian), which we, by definition of our problem, do not have. HGRec++ is based on item-user pairs for recommendation systems, making it difficult to adopt for our CICs as well. 
Gao et al. proposed using k-means with different granularities to generate a hypergraph for 3-D object recognition and retrieval~\cite{Gao2012ThreeDObjectRetrieval}. Each cluster generated through these different granularities then serves as a hyperedge. Although this hypergraph is only an initial step in their framework, it may also work for our task.

\subsection{Visualization of hypergraphs}

Traditional approaches, such as Venn or Euler diagrams, rely heavily on color and geometric shapes to differentiate hyperedges. While these methods can be effective for small hypergraphs, they quickly become illegible when the number of nodes and hyperedges increases. One strategy is to extend node-link diagrams (as used in standard graphs) with additional visual cues to represent hyperedges. An example is Bubble Sets\cite{Collins2009BubbleSets}, which draws 'bubbles' around related nodes to indicate set membership. Subsequent techniques refined the idea of region hulls for better clarity\cite{Dinkla2012KelpDiagrams,Meulemans2013KelpFusion}. Oliver et al.\cite{Oliver2024ScalableHypergraphVisualization} introduced a polygon-based visualization method using iterative simplification to reduce overlaps and enhance readability for large-scale hypergraphs. While significantly improving scalability and visual clarity over traditional layouts, their method has limitations: oversimplification can obscure key structures, and some visual clutter persists. To address these, they propose a structure-aware simplification technique that prunes and adjusts congested areas, preserving critical relationships like prominent hyperedge cycles or community connections. Their method significantly improves scalability and visual clarity for abstract hypergraphs. But in hypergraphs of image collections, where many hyperedges can overlap the same set of images, their simplification process merges away important distinctions. Even though the process can be reversed (e.g., when zooming in), the resulting visualization would still suffer from severe overlaps that make it unreadable.

MetroSets\cite{Jacobsen2021MetroSets} uses a metro map metaphor, drawing hyperedges as colored lines that run through the member nodes laid out along a schematic map. This approach is visually appealing for datasets whose sets can be meaningfully arranged as multiple paths, such as temporal progressions or other inherently ordered groupings; but it struggles with arbitrary hypergraph structures or too many overlapping hyperedges.
Others have introduced timeline- and matrix-based visualizations, which offer greater scalability. For instance, PAOHvis\cite{Valdivia2021AnalyzingDynamicHypergraphs, Valdivia2018UsingDynamicHypergraphs} uses a timeline layout to represent dynamic hypergraphs, allowing users to compare hyperedges over time. While effective for datasets with fewer than 100 hyperedges, its scalability remains limited. Matrix-based techniques, such as Hyper-Matrix\cite{Fischer2021VisualAnalyticsTemporal}, provide a more structured representation by organizing hyperedges and nodes into a grid and use semantic zooming to visualize the matrix at different scales. Set Streams\cite{Agarwal2020SetStreams} and HyperStorylines\cite{PenaAraya2021HyperStorylines} take a similar approach to visualizing hypergraphs, and, like Hyper-Matrix, are specifically aimed at visualizing temporal hypergraphs. These methods support extensive interactivity, such as filtering and hierarchical grouping, and demonstrate scaling to medium-sized datasets with several hundred elements.

In \cite{Fischer2021TowardsASurvey} it is highlighted that, despite the increasing use of hypergraphs in various domains, visualization techniques remain underdeveloped and several challenges remain. The most critical limitation is scalability: no existing technique effectively handles hypergraphs with thousands of nodes or hyperedges. This limitation stems from the visual clutter created by dense, overlapping relationships, which can quickly become overwhelming and disorienting for a user. The challenge is particularly acute for the image collections in our scope; unlike abstract nodes, each image is a visually rich item that requires significant screen space, making clutter a barrier to effective analysis and limiting the applicability of current methods.

\section{Proposed Method}
We start by considering the design criteria for our visual analytics solution, which are based on both prior literature and years of experience with complex image collections, and discussion with our co-investigators. Then we go into the representation of CICs by hypergraphs, and how to construct such a representation. Next, we introduce our interactive visualization method. Finally, we detail how constructed hypergraphs can be evaluated against a ground truth.

\subsection{Design criteria}
\label{sec:design}
Typically, experts start with unstructured collections and gradually organize them into meaningful structures through iterative exploration and targeted searches for relevant items, enabling essential insights \cite{Zahalka2014TowardsInteractive,Gisolf2021SearchExploreStrategies}. In this process, exploration is undertaken when the expert is faced with an unfamiliar collection and seeks to uncover the underlying structure through iterative dynamic analysis. In contrast, a search is used when the expert has a clear target in mind and requires a fast, precise retrieval of relevant items. Many multimedia analytics tasks, especially when analyzing CICs, require both approaches: initial exploration to understand the full scope of the data, followed by targeted searches for specific insights. Consequently, analyzing image collections involves an iterative alternation between exploration and search, and requires dynamic data exploration, filtering, searching, annotation, and hypothesis testing.

To effectively support the exploration and analysis of CICs, our method is designed around a set of criteria that reflect the specific demands of these tasks. While loosely inspired by the comparison framework proposed by Fischer et al.~\cite{Fischer2021TowardsASurvey}, our criteria focus more directly on the dual challenges of constructing and interacting with large hypergraphs derived from raw image data. Notably, Fischer et al. focus primarily on visualization techniques, whereas we additionally consider model construction, domain constraints, and the need to operate at both image and hypergraph levels.

\begin{enumerate}
    \item \textbf{Complex Image Collection Model:} The approach should be capable of representing complex, overlapping, and non-exclusive relationships among images, often present in real-world image collections;

    \item \textbf{Model Construction:} The approach should include the construction of the CIC model directly from raw image data;

    \item \textbf{Scalability:} The approach should be capable of handling image collections ranging from 1,000 to at least 100,000 images, and at least 1000 hyperedges. In particular, the hypergraph visualization must scale without inducing visual clutter.
    
    \item \textbf{Robust to Unseen Categories:} The approach should operate effectively on specialized image collections that deviate from generic datasets. It must capture not only objects but also higher-level aspects such as locations, themes, and settings, without relying on labeled training data or predefined categories.
    
    \item \textbf{Multi-Level Exploration:} The approach should support analysis at both the individual image level and the structural hypergraph level, allowing users to move fluidly between content and structure during exploration.
    
    \item \textbf{Dynamic Exploration and Search:} The approach should support an iterative analysis process by allowing users to transition between exploratory analysis and targeted search. To this end, it must provide interactive functionalities such as dynamic filtering, panning, zooming, querying and annotation, enabling users to inspect details at varying scales. Additionally, the system must maintain real-time or near-real-time responsiveness to support iterative hypothesis testing and analysis.

    \item \textbf{Orientation and Context Preservation:} To avoid cognitive overload, the approach should maintain orientation by helping users keep track of their position within large-scale hypergraphs and understand how local views relate to the overall collection, while preserving context when moving between levels of detail.
    
    \item \textbf{Usability, Interpretability, and Performance:} The interface should be designed to minimize the learning curve for domain experts by being intuitive and user-friendly, the visualization should be easy to interpret, and should work on consumer grade hardware. Together, this should result in an effective user experience.
    
\end{enumerate}

\subsection{The hypergraph model and its construction}
\label{subsec:model}
Here, we first define the hypergraph model and outline how hypergraphs are constructed from image data. We then explain how these construction methods can be evaluated, discuss existing similarity measures and their limitations, and finally introduce our own CoverEdge Similarity (CES) measure.

To support the visual exploration and search of CICs, we require a data model that captures overlapping relationships and can be derived directly from raw visual data. We adopt a hypergraph-based representation to fulfill this role. A \emph{hypergraph} is defined as an ordered pair $H = (V, E)$, where $V = \{I_1, I_2, \dots, I_n\}$ is a finite set of vertices, with each vertex $I_i$ representing an individual image in the collection. $E$ is a set of non-empty subsets of $V$; each element $e \in E$ is called a \emph{hyperedge} and represents a group of images that share common characteristics or relationships, i.e., for every edge $e \subseteq V \quad \text{and} \quad e \neq \emptyset$.

Constructing the hypergraph itself involves clustering raw image data in a way that allows for overlapping relationships. In this paper, we explore several approaches to hypergraph construction, acknowledging that the optimal method may vary depending on the specific characteristics of the image collection. Our goal is not to exhaustively determine the best method, but rather to evaluate a viable strategy that can effectively model complex image relationships. 

Pre-classification embeddings from pre-trained classification models provide an effective basis for clustering image data. Vision Transformer (ViT) models currently achieve state-of-the-art performance in image classification. Although ViTs are not specifically trained on concepts from our CICs, their extensive pre-training allows them to capture diverse visual features. In the embedding space, visually or semantically similar images naturally cluster closely, facilitating meaningful grouping. Therefore, we use pre-trained transformer embeddings for all hypergraph construction methods explored in this study.

Most pre-trained ViTs are particularly strong at recognizing objects, but they tend to be less effective at capturing thematic content, scenes, or location-related cues. This is not an inherent limitation of the ViT architecture itself, but rather a reflection of their pre-training objectives and data. To complement this limitation, we also explore several alternative models that may be better suited to these other types of features. 

To our knowledge, there are no dedicated hypergraph construction algorithms from raw data, where there are no predefined relations. We found that clustering algorithms able to assign items to multiple clusters come closest. Examples are fuzzy c-means (FCM) and possibilistic c-means (PCM). The extensive review of FCM by Askari\cite{Askari2021FuzzyCMeansClustering} shows that despite all efforts, adjustments made to FCM perform equally or worse than the baseline FCM on almost all datasets. We therefore first experiment with FCM and PCM to build our hypergraphs. We will also adopt the hypergraph construction method that is part of Gao et al.'s retrieval and recognition process~\cite{Gao2012ThreeDObjectRetrieval}.

The visualization of large hypergraphs depends fundamentally on the quality of the underlying hypergraph itself. A similarity measure is therefore essential to ensure that the visualized structures meaningfully reflect the data through the validation of the hypergraph construction methods.

In our setting, similarity measures are not used as part of the visualization process itself, but solely to evaluate and select suitable hypergraph construction methods. Constructing hypergraphs for large, unlabeled image collections cannot be directly validated without extensive manual labeling, which is infeasible. Therefore, we benchmark several construction methods on CICs where ground-truth labels are available. By comparing the constructed hypergraphs against this ground truth using a similarity measure, we can assess which method best captures meaningful structures. Once such a method is selected, it can be applied in an unsupervised way to new image collections where no labels are known.

To evaluate the quality of a constructed hypergraph, we need a measure that is able to compare (large) constructed hypergraphs with the corresponding ground truth hypergraph. For good interpretation of a visualized hypergraph, the most important qualities are a low number of images with dissimilar labels within one hyperedge (high internal precision at a local level) and no over-segmentation of the hypergraph (global level). A measure of hypergraph quality should therefore capture local and global sensitivity. A hypergraph containing many hyperedges that have low internal precision is difficult for an analyst to assess, and requires many more actions to clean up. A hyperedge with too much oversgementation may contain high precision hyperedges, but the analyst will have a much more difficult time to assess the overal content of a hypergraph, there will be increased clutter, and the analyst has to perform many additional actions to merge hyperedges to clean up the hypergraph.

Although not the most important, the measure should be computationally efficient, given the scale and complexity of hypergraphs for large image collections. Note that a similarity measure is used only when selecting and validating a suitable hypergraph construction method using annotated CICs; the chosen method can subsequently be applied to unannotated CICs.

As noted in section~\ref{sec:related_works}, the measure in \cite{Surana2023HypergraphSimilarityMeasures} is not applicable for our use case. In their paper, they also noted that to their knowledge, no other hypergraph evaluation measure existed, which was our finding at first as well. 
However, we came across an adaptation of NMI to evaluate overlapping community finding algorithms\cite{McDaid2011NormalizedMutualInformation}. These overlapping communities are nearly identical to hypergraphs, and we found that we could adopt their method for hypergraphs without making any changes. We will refer to this measure as hypergraph NMI (hNMI) in the remainder of this paper. hNMI provides a macro-level assessment by focusing on the overall similarity of the hypergraph. There is a known issue, where NMI tends to prefer solutions with more clusters because the entropy-based normalization does not properly penalize over-segmentation\cite{Amelio2016CorrectionForCloseness}, but the authors of \cite{McDaid2011NormalizedMutualInformation} note that they implemented a solution for this.

The main limitation of hNMI is that it only provides a macro-level assessment of hypergraph similarity, implicitly assuming that the importance of a hyperedge scales with its size. However, in real-life image collections, smaller hyperedges can be just as significant as larger ones. Additionally, it remains unclear whether hNMI adequately resolves the known tendency of NMI to prefer over-segmented solutions. To address these concerns, we developed a new hypergraph evaluation measure that emphasizes a micro-level perspective, treating each hyperedge equally regardless of size, and explicitly penalizing over-segmentation.

Our quality measure $S$ evaluates a generated binary hypergraph $H_g$ against a ground truth hypergraph $H_t$ by assessing how well the ground truth hyperedges are covered. First, we compute the true positive matrix with $T = H_t^T \times H_g$, where each entry $T(i,j)$ represents the number of nodes shared by the $i$-th ground truth hyperedge and the $j$-th generated hyperedge. For each ground truth hyperedge $i$ (with $\lvert H_{t_i}\rvert$ positive nodes), we define the candidate score for a generated hyperedge $j$ as
\begin{equation} 
c_{ij} = \frac{T(i,j)^2}{\lvert H_{g_j}\rvert},
\end{equation}
where $\lvert H_{g_j}\rvert$ is the size of the $j$-th generated hyperedge. Using a greedy strategy, we first select the generated hyperedge $j^*$ that maximizes $c_{ij}$ to obtain an initial score. Then, to penalize over-segmentation, additional hyperedges are added iteratively with a diminishing returns factor, dividing the candidate score by the number of hyperedges selected, until all positive nodes in $H_{t_i}$ are covered. 

The score for a given ground truth hyperedge $i$ is then:
\begin{equation} \label{eq:score_i}
S(i) \;=\; \frac{c_{i,j^*}}{\lvert H_{t_i}\rvert}
\;+\;
\sum_{k=2}^{K}
\frac{c_{i,j_k}}{\lvert H_{t_i}\rvert \,\cdot\, k},
\end{equation}
where $K$ is the total number of selected generated hyperedges required to cover all positives. The quality measure is obtained by averaging all per‐hyperedge scores:
\begin{equation} 
S \;=\; \frac{1}{p} \sum_{i=1}^{p} S(i),
\end{equation}
where $p$ is the number of ground truth hyperedges. There is one way to cheat our measure: by generating every possible combination of node/hyperedge combination, the greedy search for the best hyperedge will always return the perfect hyperedge, resulting in a perfect score. We therefore designed a diagnostic adjustment as well. Let $m$ be the total number of generated hyperedges in $H_g$, and let $u$ be the number of those hyperedges that are actually used to cover at least one ground truth hyperedge. We define the used hyperedge ratio as $R = \frac{u}{m}$. A higher value of $R$ indicates that more of the generated hyperedges were actively employed in covering $H_t$, implying fewer redundant or “cheating” hyperedges.

By multiplying $S$ and $R$, we ensure that only solutions achieving both high coverage and efficient (non-redundant) hyperedge usage score well. We get our final CoverEdge Similarity (CES) with:
\begin{equation} 
CES = R \cdot S.
\end{equation}

Within a hypergraph, it is often useful to compute local similarities. For example, to compare hyperedges \(e_a,e_b\in E\) within a constructed hypergraph $H$, we use cosine similarity between their centroid embeddings,
\(sim_{edges}(e_a,e_b)=\frac{\bar{\mathbf{x}}_{e_a}^{\top}\bar{\mathbf{x}}_{e_b}}{\lVert\bar{\mathbf{x}}_{e_a}\rVert\,\lVert\bar{\mathbf{x}}_{e_b}\rVert}\),
with \(\bar{\mathbf{x}}_{e}=\frac{1}{|e|}\sum_{I_i\in e}\mathbf{x}_i\) and \(\mathbf{x}_i\) the embedding of image \(I_i\).

Items (images) within a hyperedge can be ordered and compared using cosine similarity of their embedding vectors, \(sim_{images}(i,j)=\frac{\mathbf{x}_i^{\top}\mathbf{x}_j}{\lVert \mathbf{x}_i\rVert\,\lVert \mathbf{x}_j\rVert}\), and this same similarity can be used to rank images within the entire set of vertices $V$.

\subsection{Visualization and interaction}
Once the hypergraph has been constructed, effective exploration and interpretation depend on how it is visualized and how users can interact with it. Because exploration and search require both moving between image content and structural relationships and sometimes viewing them simultaneously, no single visualization suffices. To meet this need, our approach provides multiple coordinated views and interactive capabilities tailored to large CICs. The user can interact with the hypergraph in several ways: by visualizing it in different forms, by giving names  to  hyperedges (i.e., classify), and by modifying the hypergraph through merging, splitting, removing, or creating hyperedges, and removing or adding images to hyperedges. 

\begin{figure}[tb]
  \centering 
  \includegraphics[width=\columnwidth]{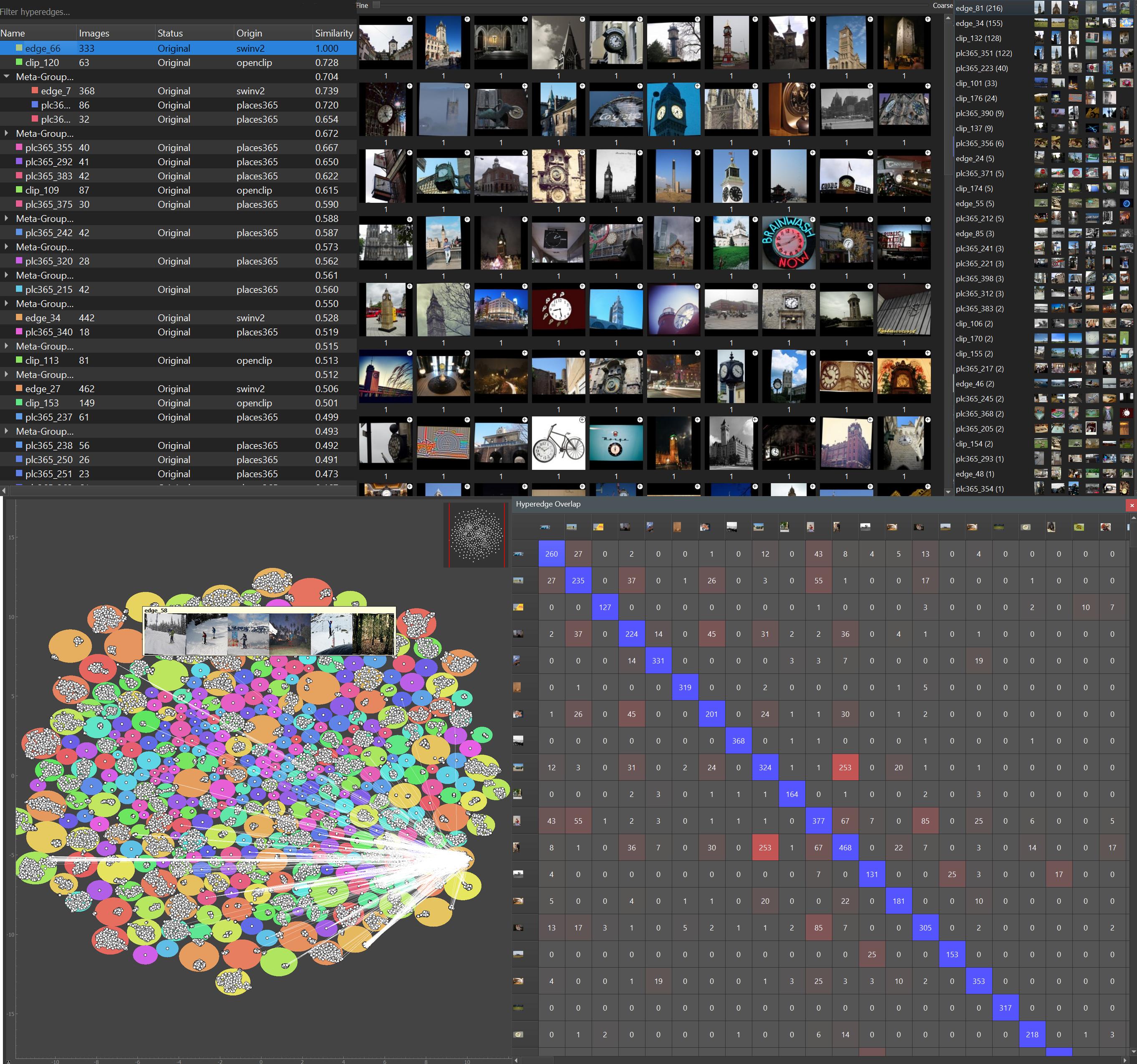}
  \caption{%
The system, with the Hyperedge list (top left), Grid visualization with intersecting hyperedges (top right), Spatial Hypergraph visualization (bottom left) and the Hypergraph matrix visualization (bottom right).%
  }
  \label{fig:app}
\end{figure}

To visualize and modify the hypergraph, the user is presented with four coordinated views of the image collection's hypergraph. The \textbf{Hyperedge list}, the \textbf{Spatial Hypergraph Visualization}, the \textbf{Grid visualization} and the \textbf{Hyperedge relation matrix visualization} (Figure \ref{fig:app}).

User interactions in one view can instantly inform and update the others, ensuring a cohesive, synchronized exploration experience that helps (7) orientation and navigation as the user navigates through large-scale image collections.

\subsubsection{Hyperedge list}
The Hyperedge list provides a comprehensive overview of all hyperedges in the current collection. Users can filter hyperedges by name and are presented with several informative columns: the number of images in each hyperedge, its status (indicating whether it has been modified by the user), its origin (e.g., embedding model, metadata, or user-generated), and the standard deviation of cosine similarity among the images within the hyperedge, relative to the hyperedge’s average embedding, which is a rough indicator of internal coherence. To aid comparative analysis, users can select a hyperedge and sort all others based on cosine similarity, with the resulting scores displayed in a dedicated column. The interface also shows the intersection size between the selected hyperedge and others, helping to identify overlap. To organize a possibly long list of hyperedges, the user can group similar hyperedges into meta-edges based on $sim_{edges}$. This affects ordering and grouping in the Hyperedge list only, not the hypergraph model. Users can adjust the similarity threshold for grouping and optionally consolidate meta-edges into single hyperedges when appropriate (this does affect the hypergraph). 

Users can add a new hyperedge or select a hyperedge to change its name, display its content, add or remove images, remove the hyperedge, or query the image collection based on the hyperedge. 

\subsubsection{Hyperedge Grid Visualization}
The Hyperedge Grid Visualization allows both quick assessment and complete understanding of a user-selected hyperedge, the results of a query, or a selection of images from one of the other views. 

The Hyperedge Grid Visualization presents a scrollable main grid of images, each displayed at 128×128 pixels (user adjustable). Users can select images from the grid to add them to a hyperedge or to use them as input for a query. Double-clicking an image opens a full-sized version, displays its metadata, and allows the user to select a region of interest for further querying. Since some hyperedges may contain hundreds or even thousands of images, this view can quickly become overwhelming. To address this, we implement hierarchical clustering (using cosine similarity over image embeddings) within each hyperedge. A slider allows the user to consolidate the displayed images into subclusters based on this hierarchy. Each subcluster can be individually expanded or collapsed, enabling a hyperedge with a thousand images to be reduced to a compact summary, such as two representative subclusters. This interaction is purely presentational and does not modify the hypergraph itself.

The Hyperedge Grid can also display an overview of all hyperedges in the hypergraph. Each hyperedge is represented by its name and six images arranged in a 2×3 layout: the top three are those most similar to the hyperedge's average feature vector, while the bottom row contains one image least similar to the average, and two images that are most dissimilar to each other. This summary provides insight into the central theme of the hyperedge, as well as its internal variation.

Adjacent to the grid is a dynamic list showing all hyperedges that contain any of the currently displayed images. Each hyperedge in this list is represented using the same six-image summary and is annotated with the number of intersecting images it shares with the current grid selection.

To help users prioritize and keep track of their interactions, images displayed on the Hyperedge Grid as part of a query result are visually marked based on their status. Images belonging to hyperedges that have been modified by the user are outlined with a colored border, while those that are part of the most recently selected hyperedge are marked with a different color. Images that have not been interacted with remain unmarked. Users can choose to hide all bordered images, allowing them to focus on previously unexamined images. This aids efficient navigation and prioritization during analysis.

\subsubsection{Spatial Hypergraph Visualization}
The Spatial Hypergraph Visualization is designed to achieve two primary goals. First, it supports analysis and exploration at the structural hypergraph level by providing users with an intuitive 2D spatial overview, where closeness between hyperedges represents similarity. Second, it supports analysis and exploration at the image relational level. Users can interactively zoom and pan the view. 

\begin{figure}[tb]
  \centering 
  \includegraphics[width=\columnwidth]{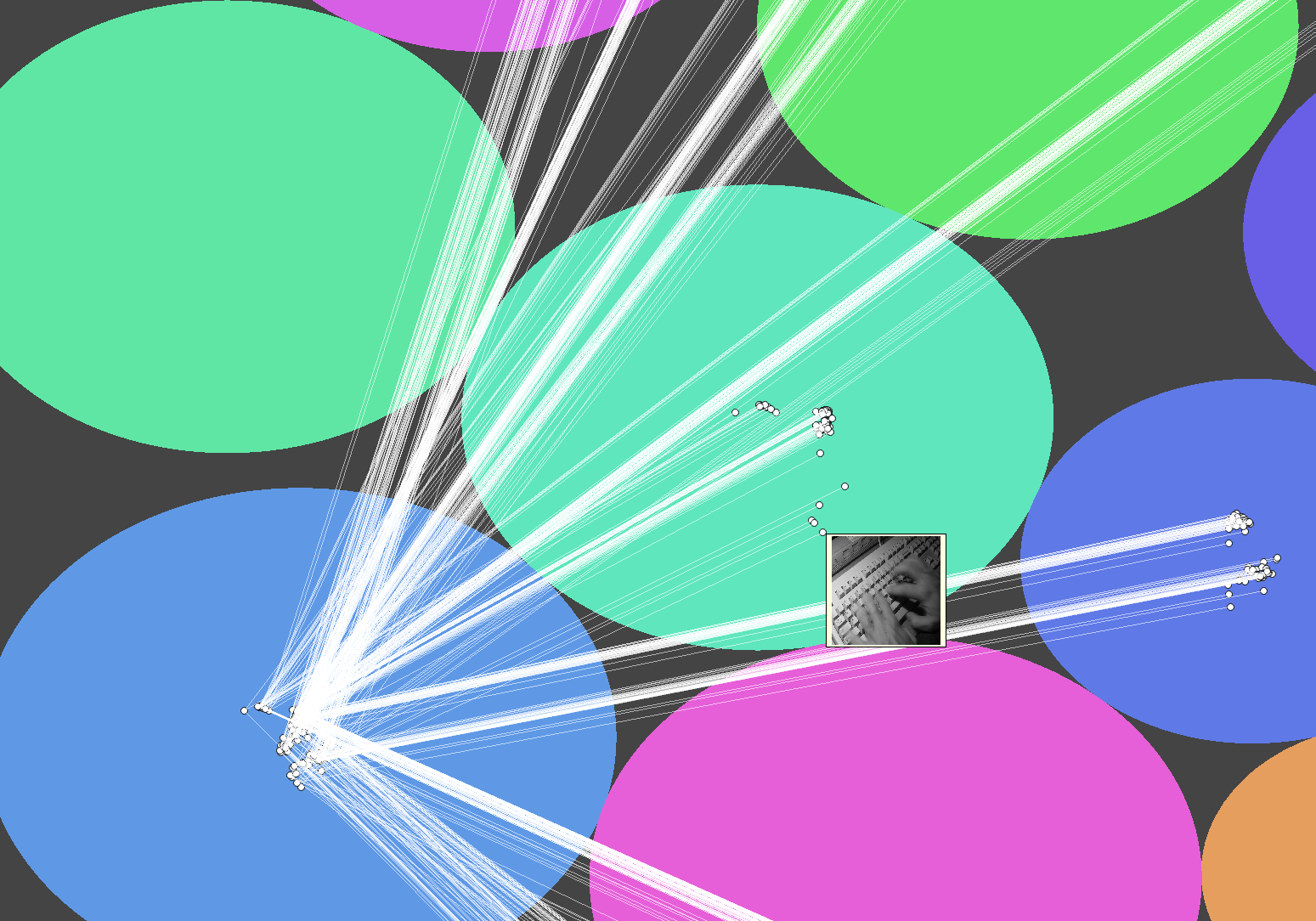}
  \caption{%
Zooming into the Spatial Hypergraph view, the user can see and select the individual image nodes. The image nodes are placed based on the internal hyperedge UMAP projection. 
  Images closer together are likely to be more similar.%
  }
  \label{fig:hg_vis_lar}
\end{figure}

For the hypergraph level, we calculate a 2D UMAP projection~\cite{lel2018umap}, using the average feature vector of all images in each hyperedge. This way, similar hyperedges are located near each other, and dissimilar hyperedges are located farther away from each other. Each hyperedge is represented by a node whose size is scaled to the number of images it contains. To prevent overlap between nodes, we apply an iterative procedure: whenever two nodes overlap, they are pushed apart until a non-overlapping layout is achieved. This ensures that larger nodes remain visible and separable, while the relative distances from the UMAP projection are preserved as much as possible. When zooming in on a specific hyperedge node, inside the node the image nodes will be shown. The image node locations within the hyperedge node are also determined by a per hyperedge generated UMAP projection (Figure~\ref{fig:hg_vis_lar}). When an image is part of multiple hyperedges, the image node for this image is present in all the hyperedges it is part of. Furthermore, a link is drawn between the image nodes that represent the same images, showing the relations between hyperedges and images. To prevent clutter, only the hyperedge or individual images selected by the user are shown in this fashion, as well as the hyperedges that intersect with the selected hyperedge.

Hovering over a hyperedge node displays a tooltip with the six-image summary, and hovering over an individual image node gives a tooltip with a thumbnail preview of the image. Users can use a lasso tool to select multiple image or hyperedge nodes, for which the links are then highlighted in a different color, and the images are displayed on the Hyperedge Grid. 

To help users locate specific hyperedges from the Hyperedge List within the Spatial Hypergraph View, we implemented dynamic highlighting. When a user selects a hyperedge in the Hyperedge List, its corresponding node(s) in the Spatial Hypergraph View are emphasized with a short pulsing animation and a distinct highlight color. Similarly, when one or more images are selected in the Hyperedge Grid, all hyperedges containing those images are highlighted in the same manner, and their connecting links are rendered in a distinct color to further aid identification.

\subsubsection{Hypergraph Matrix Visualization}
The Hypergraph Matrix Visualization provides a quantitative overview of relationships between hyperedges, supporting multi-level exploration at the structural hypergraph level. Each cell indicates the number of images shared between pairs of hyperedges, offering a concise summary of overlaps. Cell colors encode the harmonic mean of overlap, ranging from dark grey (minimal) to red (maximal). Each hyperedge is accompanied by a representative image, allowing quick visual assessment of its contents. Hovering over a cell shows a tooltip with the six-image summary for each of the respective hyperedges. Users can select a cell to display the intersecting images of these hyperedges on the Hyperedge Grid, isolating specific combinations, such as a person and a location. As the matrix only visualizes hyperedges, it remains scalable even for large image collections. Users can zoom out to get a higher level overview.

\subsubsection{Cross view interactions and hypergraph model interactions}
To help users maintain an overview of progress and status, hyperedge node colors can be switched to reflect state (e.g., modified, new, original) or origin (e.g., user-created, model-based), and in all cases the node colors are kept synchronized with the corresponding entries in the Hyperedge list. 

Selecting images or hyperedges in one view, highlights these in the other views, and users can double click an image to view the full size original, and metadata (if available). Any modification to the hypergraph, such as merging, splitting, renaming, or reassigning images to hyperedges, instantly updates across all views. This ensures that the user always works with a consistent view of the data.

We implemented several methods for querying the image collection:
\begin{itemize}
    \item \textbf{Image query:} Uses the average feature vector of one or more selected images.    
    \item \textbf{Hyperedge query:} Uses the average feature vector of a selected hyperedge.    
    \item \textbf{ROI query:} Uses the feature vector of a user-selected region within an image.    
    \item \textbf{Clipboard query:} Uses the feature vector of an image from the operating system clipboard.
    \item \textbf{Text query:} Uses an OpenCLIP embedding of a textual description. 
\end{itemize}

The queries can be used to find additional images to add to a specific hyperedge, or to create entirely new hyperedges. The text query specifically is not just useful for finding specific objects, but also works well for more abstract, thematic queries. Each query makes use of $sim_{images}$.

While the constructed hypergraph is based on visual similarity, metadata can offer an alternative perspective on the image collection. To support this, users are provided with tools to generate hyperedges based on metadata. A dedicated interface displays all available EXIF metadata fields present in the collection, along with an overview showing the number of images that contain a valid (non-empty) value for each field and the number of unique values it contains. With a single action, the user can add a metadata field to the hypergraph. This results in the creation of a general hyperedge containing all images with a valid value for that field, as well as separate hyperedges for each unique value (in the case of categorical data), or a binned range (in the case of continuous values), each containing the corresponding subset of images. These hyperedges are automatically named based on the metadata field and value, and are labeled as originating from "metadata" in the Hyperedge List for easy identification. In the Spatial Hypergraph View, metadata-based hyperedges are positioned along the right edge to distinguish them from visually derived hyperedges.

To make our UI as accessible as possible, both for new and for infrequent users, we use text buttons with a clear description of their function as opposed to icons~\cite{Wiedenbeck1999UseIconsLabels}.

\begin{figure}[!t]
  \centering 
  \includegraphics[width=\columnwidth]{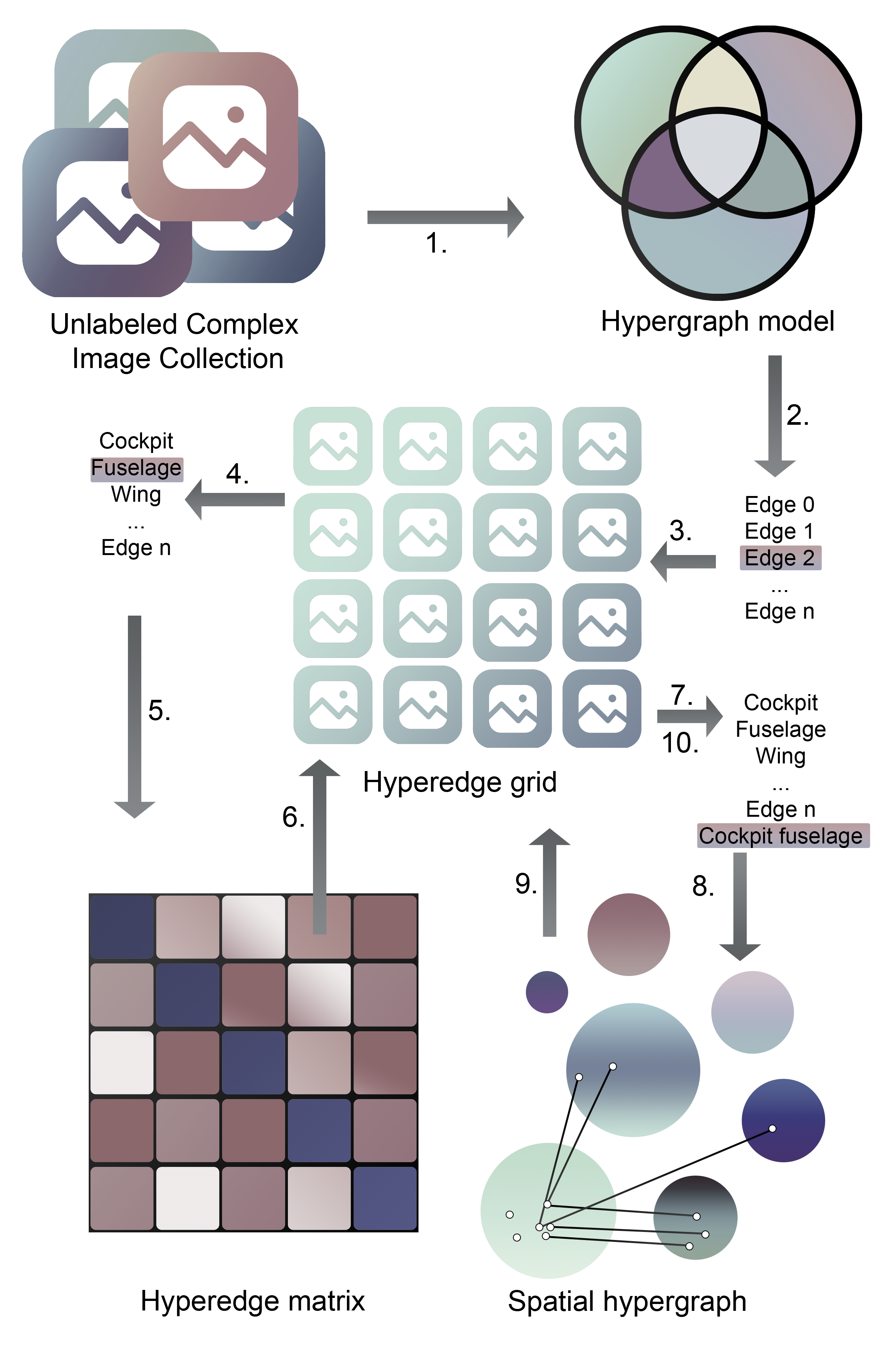}
  \caption{
  	Typical workflow with our method for a CIC (in this case, MIC). The expert would first construct the hypergraph (1), then select (2) and visualize a hyperedge from the Hyperedge list using the Hyperedge Grid (3) and give it a label based on its content (4). Once the expert has given some structure by labeling, the Hypergraph matrix can be used to see if there are any interesting overlaps (5) between the cockpit and the fuselage hyperedge. The intersecting images can then be displayed (6) and the expert may see there are indeed images that depict the fuselage of the cockpit specifically. The expert can then decide to make a new hyperedge for this (7) and visualize the new hyperedge on the Spatial Hypergraph (8) to find additional candidates for this hyperedge. Using the lasso tool (9) these can be displayed on the grid and any relevant images can be added to the hyperedge (10).
  }
  \label{fig:schema}
\end{figure}

Figure~\ref{fig:schema} shows how an expert could use our approach to analyze a complex image collection.

\section{Results and evaluation}
\label{sec:results}
In this section, we first validate the hypergraph similarity measures, CES and hNMI, which we need to evaluate which construction methods yield useful hypergraphs. We then describe the complex image collections used as evaluation benchmarks. Next, we present the results of the hypergraph construction algorithm evaluation. Finally, we report on user evaluation sessions with domain experts to assess whether the designed system, including the hypergraph generated by the selected hypergraph construction algorithm, is useful in practice and to understand how it is employed in real investigative workflows.

\subsection{Hypergraph similarity measure validation}
As noted in section \ref{sec:design} a high quality hypergraph that is optimal for visualization contains hyperedges with high internal precision and no over-segmentation. To assess whether CES and hNMI are able to distinguish low quality hypergraphs from high quality hypergraphs, we performed several experiments with synthetic hypergraphs. For comparison, we also implemented the Tensor-Hamming and Tensor-Centrality measures from \cite{Surana2023HypergraphSimilarityMeasures}. Although Tensor-Spectral (Tensor-H) is their best performing measure, due to proprietary code we were not able to implement it, and due to limitations of these measures, not all experiments could be performed with the tensor measures. Since the validation of the hypergraph similarity measure is not the main focus of the paper, results for the tensor methods, as well as more detailed results can be found in the supplemental materials.

We first test the robustness of the similarity measures by increasingly perturbing the hyperedges of a ground truth hypergraph and then comparing it to the initial ground truth hypergraph. A good similarity measure should show a monotonically decreasing similarity score with increasing perturbation. This shows us whether a measure is able to distinguish low precision hyperedges from high precision hyperedges.

The initial ground truth hypergraph is created using the Erd\H{o}s--R\'enyi (ER) hypergraph constructor\cite{Erdos1959OnRandomGraphs}. We perturb this ground truth hypergraph in two ways. First, we replace a percentage of hyperedges with hyperedges with the same number, but randomly chosen vertices. Second, for a percentage of hyperedges, we replace a vertex with a random other vertex not already in the hyperedge. Both experiments give us similar results, with hNMI following the perturbation percentage almost perfectly and CES deviating only a little.  

As noted in subsection \ref{subsec:model}, hNMI and CES in theory differ in macro versus micro level. To see if this is the case in practice, we measure the effect of perturbing small versus large hyperedges. We generated synthetic ground truth hypergraphs that consist of half large (100 vertices) and half small (10 vertices) hyperedges. We then perform two experiments. One where we perturb only the small hyperedges, and one where we perturb only the large hyperedges. This indeed informs us that hNMI is not affected as much when only smaller hyperedges are perturbed, whereas CES is, and vice versa. Neither is wrong and is based on preference and goal of the measurement. CES can be adjusted to behave like hNMI by using a weighted averaging per hyperedge, based on the number of vertices. 

Finally, we set up an experiment to see in what way CES and hNMI react to over-segmentation. We increasingly subdivide the hyperedges of a synthetic hypergraph, while keeping the parent edges as well. We compare the perturbed hypergraph to the ground truth hypergraph. hNMI does not show problems with oversegmentation, nor does CES. As intended, CES punishes additional redundant hyperedges more.

Both hNMI and CES are capable of distinguishing lower from higher quality hypergraphs. hNMI tends to follow perturbation levels more consistently, while CES deviates slightly. Unlike hNMI, CES can be tuned to emphasize either small or large hyperedges depending on the evaluation goal. As they capture different aspects of hypergraph quality at low computational cost, using both can be informative, though either measure alone already provides reliable results. This allows us to properly evaluate hypergraph constructors and identify those suitable for visual analytics. The tensor-based measures, by contrast, either fail to capture degradation consistently or (as expected) do not scale to larger hypergraphs, making them less practical for our purposes.

For completeness, we also compared hNMI and CES against the tensor-based methods on synthetic hypergraphs generated by the Erdős–Rényi, Barabási–Albert, and Watts–Strogatz models, following the evaluation protocol of \cite{Surana2023HypergraphSimilarityMeasures}. Detailed results, including ROC and UMAP visualizations and comparisons with Tensor-Hamming and Tensor-Centrality, are provided in the supplementary materials. Both hNMI and CES performed well at distinguishing between these generative models, with CES showing the strongest separation and hNMI close behind. In their study, the Tensor-Spectral (Tensor-H) method was reported as their strongest performer, achieving results comparable to what we observe for CES, but we were unable to evaluate it directly due to unavailable code.

\subsection{Image collections}
To evaluate our method, we seek multi-label image collections that vary in visual similarity and category diversity while also resembling the type of image collection encountered in real-life investigations. Public datasets fall short in this respect, so we complement them with confidential collections from safety investigations to ensure the evaluation reflects realistic investigative scenarios.

Table~\ref{tab:datasets} provides an overview of the datasets used in our evaluation.  
\begin{table}[t]
\centering
\small
\caption{Overview of the evaluation datasets.}
\resizebox{\columnwidth}{!}{%
\begin{tabular}{@{}llrrl@{}}
\toprule
Dataset & Domain & Images & Labels & Public \\ \midrule
CUB-200~\cite{CUB200} & Birds & 11,788 & 512 & Yes \\
MLRSNet~\cite{MLRS} & Remote sensing & 109,161 & 60 & Yes \\
DSEG660~\cite{DSEG} & General (COCO-like) & 97,774 & 80 & Yes \\
MIC~\cite{MH17} & Airplane crash & 12,000 & 27 & No \\
MAIC & Maritime accidents & 42,960 & – & No \\
PAIC & Parking garage accident & 4,417 & – & No \\ 
\bottomrule
\end{tabular}%
}
\label{tab:datasets}
\end{table}

\begin{figure}[!hbtp]
  \centering 
  \includegraphics[width=\columnwidth]{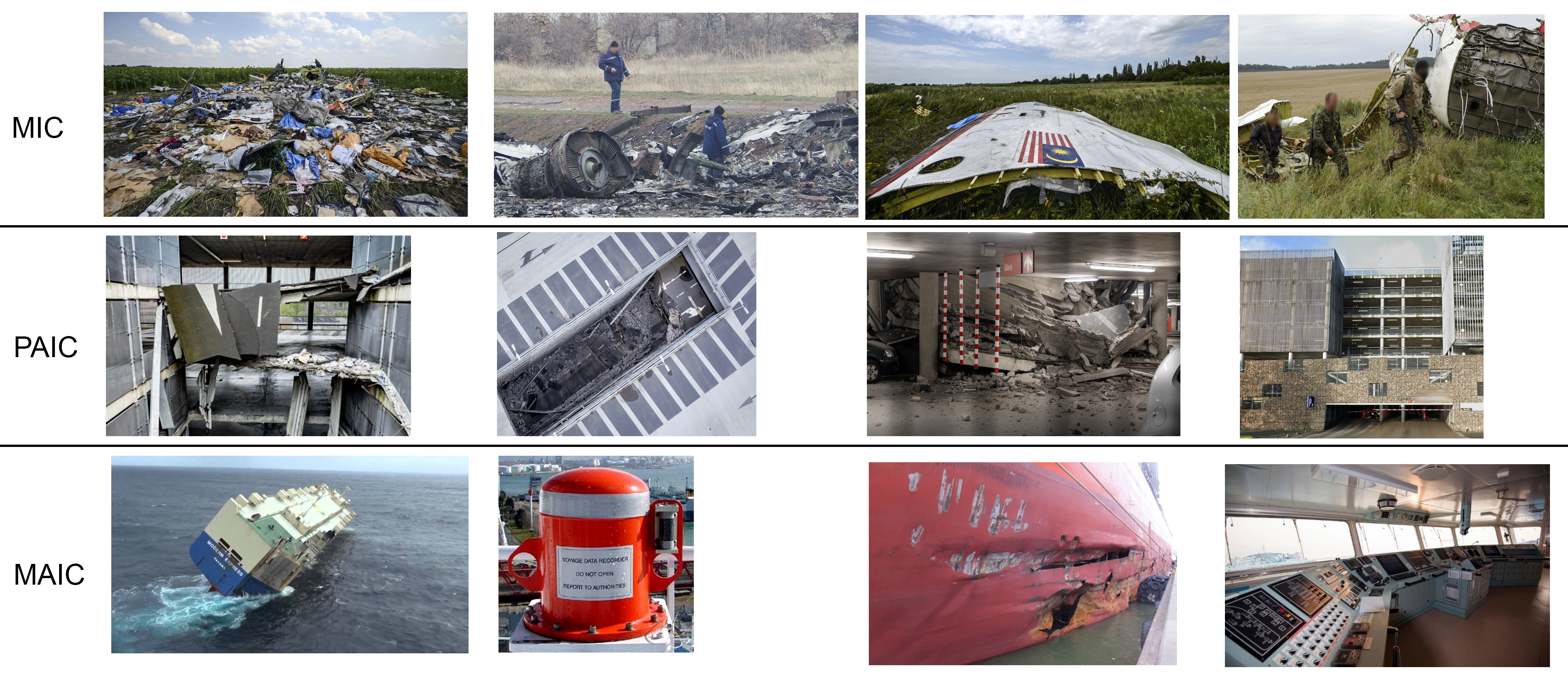}
  \caption{%
  	Some examples of the images in the accident investigation image collections.%
  }
  \label{fig:examples}
\end{figure}

CUB200, MLRS, DSEG660 and MIC are used to evaluate hypergraph construction methods. Only the first author, part of the DSB, has had full access to the classified image collections. Co-authors have seen examples of content. Example images shown are from publicly released images. The MH17 image collection (MIC), Marine accident image collection (MAIC) and Parking garage accident image collection (PAIC) are used for the domain expert user evaluation sessions and are image collections used during actual accident investigations. Figure \ref{fig:examples} shows samples of these collections. For the other collections we refer to their respective references.

\subsection{Hypergraph construction evaluation}

To evaluate the hypergraph construction algorithms, we first extract feature vectors from the images. Our primary feature extractor is a Swin v2 model pre-trained on ImageNet1k~\cite{swinv2}, a state-of-the-art architecture for image classification. Because Swin v2 is primarily trained for object recognition, we also explore incorporating features from two additional pre-trained models: Places365 and OpenCLIP. Places365 is trained specifically for scene and location recognition, which could provide embeddings that emphasize environmental context. OpenCLIP, a contrastive vision-language model trained on detailed image-caption pairs rather than single-label annotations, may capture multi-label semantics and broader scene-level information. Both models are also integrated into the hypergraph visualization system, with OpenCLIP additionally enabling text-to-image retrieval. Unless stated otherwise, all reported experiments use only Swin v2 embeddings.

We use the extracted embeddings as input to the hypergraph construction algorithms. FCM produces a membership degree for each image to every hyperedge (cluster), which we convert into binary assignments using a threshold $t$. It has two key parameters: the number of clusters $k$ and the fuzzifier $f$, which controls clustering softness; lower values yield more distinct clusters, while higher values increase membership overlaps.

\begin{figure*}[ht]
       \includegraphics[width=\textwidth]{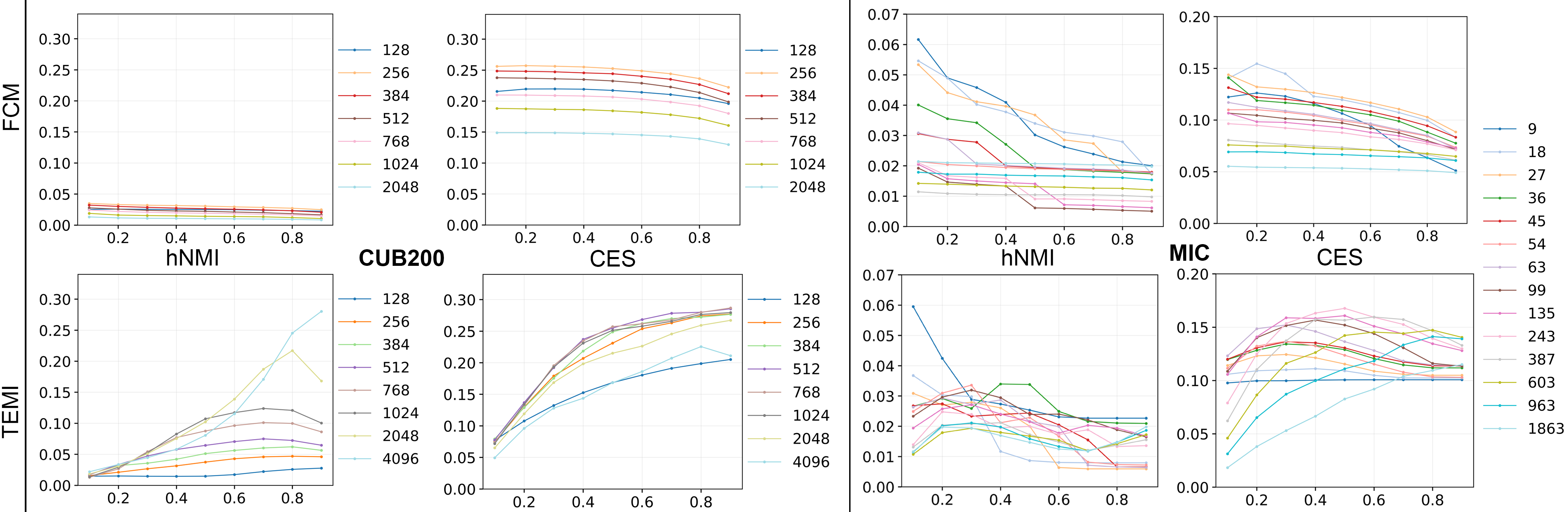}
          \caption{Results for various combinations of $t$ and $k$ for CUB200 and MIC, using either FCM or TEMI to construct the hypergraph and hNMI or CES to evaluate the constructed hypergraph against a ground truth hypergraph. The results show that TEMI outperforms FCM and that CES is more robust to over-segmentation than hNMI. FCM did not finish with $k=4096$}
     \label{fig:allresults}
   \end{figure*}

We observed that values of $f>1.1$ resulted in assigning equal probabilities to all hyperedges. Conversely, very low values of $f$ result in minimal overlap, effectively reducing FCM to standard (hard) clustering, as confirmed by experiments on the CUB200 image collection (figure~\ref{fig:allresults}), where results were mostly unaffected by $t$. We also found that FCM computation time and memory consumption are very high when the number of hyperedges increases to beyond 50 or the number of images were more than 50 000. While the computation time of the generation of the initial hypergraph may not have to be a major concern, this does depend on how easily suitable parameters can be identified to construct a meaningful hypergraph.

We tried PCM with several combinations of parameters, which are similar to FCM. However, PCM did not result in hypergraphs that performed much better than chance. Similarly, we tried the method by Gao et al.~\cite{Gao2012ThreeDObjectRetrieval} with different variations of sets of $k$, but were not able to produce any useful hypergraphs. 

Given the disappointing results from existing methods, we turn to state-of-the-art clustering algorithms designed for single-label data. To select a useful candidate for constructing a hypergraph in our context, it needs to be easy to adapt to multi-label clustering and computation time should not be excessive.  

We found Teacher Ensemble-weighted pointwise Mutual Information (TEMI), currently among the top performing clustering algorithms for single-label datasets. For a full formal definition, see ~\cite{Adaloglou2023ExploringDeepImageClustering}, here we summarize the architecture and training objective to provide enough context for understanding our integration of TEMI into the hypergraph construction pipeline. TEMI is a multi-stage, self-distillation framework for unsupervised image clustering that builds on a pre-trained image classification model. In the first stage, feature vectors are extracted from images with the pre-trained model. Clustering is then approached by assuming that an image and its nearest neighbors in this feature space are likely to share a semantic label. To address the inherent noise in this assumption, since some neighbors may be semantically unrelated, TEMI trains a student clustering head to predict a user-specified number of cluster assignments, guided by a teacher head updated as an exponential moving average of the student. The framework uses a loss based on pointwise mutual information, with instance-level weighting derived from teacher predictions. This weighting mechanism reduces the influence of inconsistent or ambiguous image pairs.

Rather than relying solely on individual pairs, the method learns cluster structure from broader neighborhood consensus: consistent patterns across many nearby samples. This is further reinforced by averaging predictions over an ensemble of clustering heads. The method achieves strong performance on standard benchmarks without fine-tuning the pre-trained model. 

Although TEMI includes a training phase for self-distillation of clustering heads, its computational cost remains relatively low compared to FCM. This makes it practical for real-world applications. Furthermore, TEMI originally assigns images to the cluster with the highest membership probability. This makes it easy to modify by applying a threshold instead of assigning it to the single highest probability cluster. This allows images to belong to multiple clusters (hyperedges). This means TEMI requires two user-defined parameters: a threshold and the number of clusters.

In Figure~\ref{fig:allresults} we show the results for the constructed hypergraphs by FCM and TEMI on each image collection, compared to their ground truth. The adapted TEMI clustering method produced better results and outperformed FCM for all of the tested image collections. We found that in at least one case (CUB200 with TEMI) that hNMI shows some signs that it may not always be robust against over-segmentation. CES shows no problems with over-segmentation, and CES results seem to suggest that for most image collections (except CUB200) there is a clear best value $t=0.4$. 

Since hNMI and CES identify in some cases different parameter settings as optimal, we also conducted a qualitative assessment of those resulting hypergraphs. Our evaluation focused on whether the images within each hyperedge form coherent groups, and whether intersections between hyperedges provide meaningful connections.

For the MIC, using FCM with 9 hyperedges and a threshold of $t=0.1$ produced very broad hyperedges, with only two showing a clear thematic focus. While such broad groupings are not necessarily problematic, they rely heavily on the utility of hyperedge intersections. In this configuration, we found 10 out of 73 intersections useful, enabling straightforward creation of more specific hyperedges through the Hypergraph Matrix.

Increasing the number of FCM hyperedges to 253 ($t=0.1$, the CES-optimal setting) resulted in high-precision groupings where most images within a hyperedge were visually similar. However, intersections between hyperedges were almost entirely absent, making the results similar to those of standard k-means clustering with single assignments. This lack of overlap often split conceptually related images, such as different views of the same object or location, across separate hyperedges without shared images, meaning that no links appeared in the Spatial Hypergraph view. In some cases, related images could still be located using the similarity-based sorting in the Hyperedge List.

In contrast, TEMI with 253 hyperedges and a threshold of $t=0.5$ (the CES-optimal setting) produced more overlap between hyperedges in ways that were useful, for example linking hyperedges of the same location but containing different objects. While this resulted in slightly less precise individual hyperedges compared to FCM, we would argue that the increased connectivity between hyperedges is a worthwhile trade-off. TEMI could also be tuned to produce results similar to the more distinct clusters of FCM by increasing the threshold, whereas we were unable to adjust FCM parameters to replicate TEMI’s overlapping structure.

For the CUB200 dataset, hNMI consistently rated FCM results much lower than TEMI results. Using FCM with 243 hyperedges and a threshold of $t=0.4$ (CES-optimal), we observed high-precision hyperedges but again almost no overlap. TEMI with the same number of hyperedges and threshold produced similarly precise hyperedges, but with more meaningful overlaps. This made it, for example, much easier to find all birds afloat on water: selecting one such hyperedge would link to others containing similar images.
Results for MLRS and DSEG660 were similar to those of MIC and can be found in the supplemental materials.

We evaluated OpenCLIP and Places365 embeddings by generating hypergraphs using the best parameter combinations for each image collection, as determined by CES. Because CES also provides scores for individual ground truth hyperedges, we compared these results against those obtained with Swin v2 embeddings to assess whether the alternative models could offer an advantage. OpenCLIP outperformed Swin v2 on roughly 10\% of hyperedges, and only by a small margin, while Swin v2 occasionally achieved substantially higher scores. Places365 did not outperform Swin v2 on any hyperedge in any dataset. Although Places365 is an older model, we had anticipated that its specific fine-tuning for location recognition might provide benefits in certain cases, but this was not observed. No consistent pattern emerged in the cases where OpenCLIP surpassed Swin v2. Despite the limited quantitative gains, we still include OpenCLIP in our system because it supports text-based querying.

\subsection{User evaluation}
We performed an evaluation session for our visualization method involving 8 domain experts from the Dutch Safety Board. These participants are professional accident investigators in the construction and fire, maritime, and aviation domains, who take part in every aspect of an investigation, such as interviewing those involved, writing reports and analyzing all available data. Importantly, this also means that they sometimes analyze complex image collections. Before the session, participants were informed of the study's purpose and provided their consent to take part. We used think-aloud protocol, a common technique to better understand the participants' thoughts and actions.

We tried to design our tasks such that they would require either exploration or search, or a combination. The specific tasks, as well as detailed descriptions of how the participants approached the tasks, can be found in the supplemental materials.

Across the three collections (MAIC, PAIC, MIC), participants converged on two effective entry points: the Overview for quick sensemaking and bootstrapping, and text search for fast retrieval. Most tasks were solvable with standard retrieval alone; domain knowledge amplified text queries (e.g., ship types, recognizing interior of cockpit), while ROI/Image queries consistently improved precision for fine-grained cues (e.g., isolating hull gashes or beam joints). Hyperedge-centric workflows, starting with collecting a seed set, then querying that hyperedge, scaled quickly (P1 surfaced 89 hull-damage images in 5 minutes), and noise detection (finding irrelevant items) benefited from targeted keywords (P5 surfaced 654 non-MH17 images in 10 minutes), but also from querying unique looking irrelevant items (P8 found 1955 images). The Spatial Hypergraph View was most valuable for discovering adjacent or overlapping clusters and for "where next?" navigation, rather than as a universal first step.

Where tasks required subtle disambiguation or multi-label constraints (bridge of a fishing vessel; floor-specific garage views), performance hinged on query formulation and verification strategies: times ranged from only a few seconds to ~18 minutes after multiple text-query iterations, with one case solved in under 10 seconds by pasting the full task prompt as a text query. Through spatial exploration the same task was solved in 3 minutes. 

One participant used the Intersection list to successfully find additional images. The Hypergraph Matrix was not used by the participants, except for one task, which was designed specifically to push participants towards using it, where it worked adequately. 

The participants had some quality-of-life suggestions, such as a good way to go back to a previous view of the data, when diving into query results repeatedly. An especially useful suggestion was to use the colorization of the hyperedge nodes in the Spatial Hypergraph view to indicate which hyperedges the user has seen and how long ago this was (darker color based on time since last visit). Both these suggestions were implemented after P1-P3 completed their tasks, before the next participants. Especially the navigation back and forth was frequently used by subsequent participants. 

In general, the participants were positive about the experience and about the capabilities of the system. They noted that if they encountered a large image collection, they would certainly want to try the system. Some said they were currently working on cases where the system would be relevant, so we are planning to set up the system for their data shortly after the user evaluations.

Overall, the system supports complementary strategies: The Overview for initial orientation and exploration, text and image queries for breadth and depth once they had a specific target in mind, ROI query for extra precision, with the Spatial Hypergraph View serving as a quick way to identify related hyperedges and find additional images.

\section{Discussion}
In this work, we aimed to address challenges associated with applying hypergraphs to complex, real-world image collections: constructing hypergraphs from raw, unlabeled data, evaluating their quality against meaningful ground truths, and developing a scalable, interpretable visualization. Here, we discuss how effectively our methods addressed these specific challenges, as well as the limitations and considerations identified through our evaluations.

We found a lack of research on the construction of hypergraphs from raw data. We tested several approaches to constructing hypergraphs and found that adapting TEMI allowed us to generate hypergraphs without relying on extensive labeled data or predefined categories. While this method was generally effective at uncovering latent image relationships, during the setup of the user evaluation datasets we observed that it sometimes produced hypergraphs with limited overlap between hyperedges. While we could solve this by adjusting parameters such as the number of hyperedges or threshold $t$, optimal parameter selection was collection-dependent. A possible solution would be to perform a grid search, until a certain overlap between hyperedges is achieved, but that may result in unwanted hypergraphs for image collections inherently containing images dominated by single, discrete objects with minimal natural overlap. Future work could explore more specialized hypergraph construction methods explicitly designed to accommodate varying degrees of overlap inherent in different image collections. 

To do the analysis of the hypergraph construction method, we developed a new measure that allows us to compare complex hypergraphs to ground truth hypergraphs. Based on experiments with synthetic hypergraphs, we found that CES is both sensitive to perturbations and robust against over-segmentation. hNMI showed greater stability in some perturbation tests, but our measure provided finer-grained insight into how well individual hyperedges were captured. Together, these results indicate that CES and hNMI can capture very similar aspects of hypergraph quality, but also more complementary aspects. Due to the low computation cost, they can both be used for a more nuanced evaluation of construction methods.

The challenge of scalable visualization, which has been widely acknowledged as problematic for large hypergraphs~\cite{Fischer2021TowardsASurvey}, also significantly shaped our research. Our evaluation highlights that the system is both technically scalable and usable on consumer hardware. However, scalability in visualization is not solely a matter of rendering performance. Cognitive scalability is a different challenge: large, overlapping hypergraphs can easily exceed what users can process and interpret in a single view. Our approach mitigates this through multiple coordinated views and progressive detail-on-demand. Our evaluation was conducted with domain experts who were not specialists in image analysis, but the participants were able to perform the tasks to a high degree, and did not get lost in the large amount of data they needed to sift through in a short time. This highlights the accessibility of our system. It would be interesting for future work to understand how users manage complexity over extended sessions lasting several weeks or months in real cases, how the hypergraph model is being shaped through their progressive analyses, and if the system requires additional features to support this long-term workflow. 

While the users often used standard image retrieval approaches to find relevant results, the Spatial view was also used a lot, and was found to be easy to use and navigate. The Hypergraph Matrix was not used often during user evaluations. In our own experience of using the application for actual case work, we found the Hypergraph matrix to be more useful in later stages of an investigation, when a significant number of hyperedges have already been created and/or properly labeled, something that is more difficult to simulate in a shorter user evaluation.

Another question pertains to the interpretability of the automatically constructed hypergraphs. This concerns both the coherence of the individual hyperedges, but also the ability to reveal meaningful connections between hyperedges. Participants found that some hyperedges had a clear theme, while others were more opaque, limiting their analytic value. However, they also noted this was not really an issue, as long as there was a good way to find a seed image. The successful usage of the Spatial view and the Intersection list by several participants shows there were useful relations between the hyperedges. 

Overall, the user evaluations revealed that participants employed diverse strategies to approach the tasks, yet the system was flexible enough to support each of them in successfully completing their analyses. While there remains considerable room for refinement, particularly in terms of quality-of-life features, all participants emphasized that the system was intuitive to use and aligned well with their workflows.

\section{Conclusion}
This study introduced an end-to-end approach for representing and analyzing complex image collections using hypergraphs. We addressed three central challenges: (1) constructing hypergraphs directly from raw, unlabeled data, (2) evaluating the quality of these hypergraphs against ground truth, and (3) enabling scalable and interpretable visualization for domain experts.

For hypergraph construction, we adapted the TEMI clustering algorithm to support multi-cluster assignments, producing overlapping hyperedges that better reflect the inherent structure of complex image collections. While effective across multiple datasets, results showed that optimal parameter settings remain collection-dependent, highlighting the need for future methods that more directly accommodate varying degrees of overlap.

To evaluate construction methods, we introduced the CoverEdge Similarity (CES) measure. Experiments with synthetic and real-world data demonstrated that CES is sensitive to perturbations and robust against over-segmentation, providing complementary insights to hNMI, which showed greater stability in some perturbation tests, but only provides a macro-level quality measurement. Together, the two measures enable a more nuanced evaluation of hypergraph quality at low computational cost.

For visualization, we developed a system that combines multiple coordinated views to support both exploration and search. Our evaluation with domain experts demonstrated that the system is technically scalable and usable on consumer-grade hardware, and that participants were able to complete challenging analytic tasks despite the scale and complexity of the data. Users used various strategies, and, although some views were more frequently adopted than others, the system proved to be flexible enough to support different workflows.

Overall, our findings show that hypergraph-based visual analytics can make large and complex image collections interpretable and actionable for domain experts. Future work should focus on more adaptive hypergraph construction methods, deeper evaluation of long-term analyses of complex image collections, and design refinements that further improve usability in real investigative settings.

\bibliographystyle{abbrv-doi-hyperref}

\bibliography{visbib}

\clearpage
\section*{Supplemental Material}  

\setcounter{section}{0}
\setcounter{subsection}{0}
\renewcommand{\thesection}{A.\arabic{section}}
\renewcommand{\thesubsection}{A.\arabic{section}.\arabic{subsection}}

\setcounter{figure}{0}\renewcommand{\thefigure}{A\arabic{figure}}
\setcounter{table}{0}\renewcommand{\thetable}{A\arabic{table}}
\setcounter{equation}{0}\renewcommand{\theequation}{A\arabic{equation}}


\section{Methods and results for hyperedge similarity validation}

Throughout the experiments we use several methods to generate synthetic hypergraph\cite{Surana2023HypergraphSimilarityMeasures}. For details on these methods, we refer to their paper. In short, each generator produces random hypergraphs according to specific probabilistic rules, resulting in characteristic structural patterns: 
\begin{enumerate}
\item \textbf{Erd\H{o}s--R\'enyi (ER) hypergraph.} Constructed by uniformly sampling $m$ hyperedges of size $k$ from the $\binom{n}{k}$ possible hyperedges. Produces a homogeneous hypergraph with narrow degree distribution.  
\item \textbf{Barab\'asi--Albert / Scale-Free (SF) hypergraph.} In this model, new edges are more likely to connect to vertices that already have a high degree of connections. Hyperedges are formed by sampling $k$ vertices with probabilities proportional to their rank, generating a power-law degree distribution with hub vertices.  
\item \textbf{Watts--Strogatz (WS) hypergraph.} Starts from a regular $k$-uniform hypergraph and rewires each hyperedge with probability $p$ to a random hyperedge. Produces small-world (locally dense, short path lengths) structures.  
\end{enumerate}

For the perturbation experiments, each experiment was ran 50 times, generating 50 different synthetic ground truth hypergraph with 99 nodes, 50 hyperedges and a cardinality $c$ of 4, created with the Erd\H{o}s--R\'enyi (ER) hypergraph constructor. This way it fits within our GPU's memory, and we can perform the experiments with two of the tensor based measures as well.

\begin{figure}[tb]
  \centering 
  \includegraphics[width=\columnwidth]{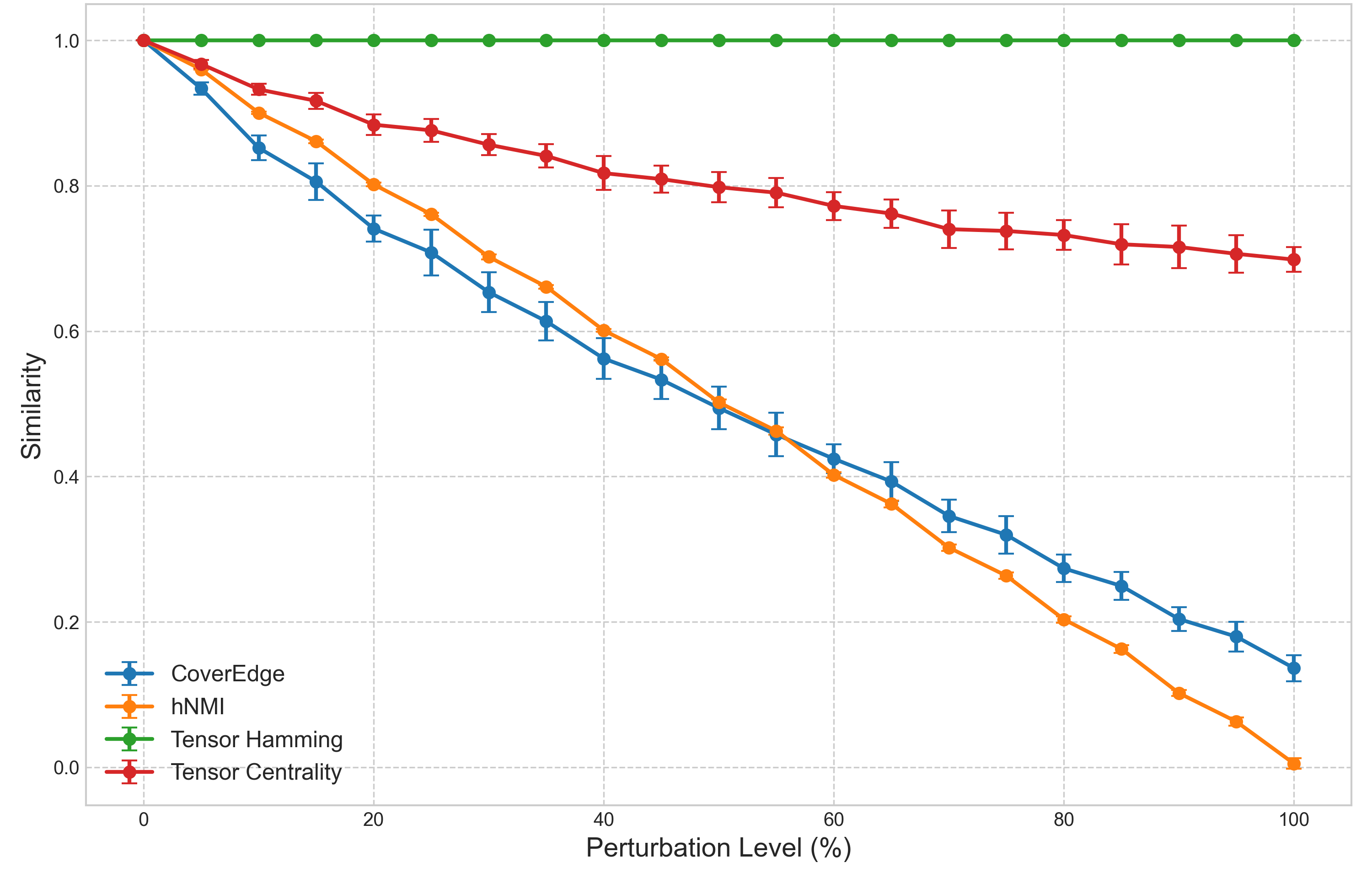}
  \caption{%
The effects of replacing correct hyperedges by hyperedges with random vertices on each measure.%
  }
  \label{fig:replace}
\end{figure}

For the first experiment, a fraction $p$ of all hyperedges is removed and replaced by entirely new hyperedges of the same cardinality $c$. These new hyperedges are generated by uniformly sampling $c$ distinct vertices from the vertex set of size $n$. This ensures that the overall number and size distribution of hyperedges remain unchanged, while their specific composition is randomized. Figure~\ref{fig:replace} shows the results.

\begin{figure}[tb]
  \centering 
  \includegraphics[width=\columnwidth]{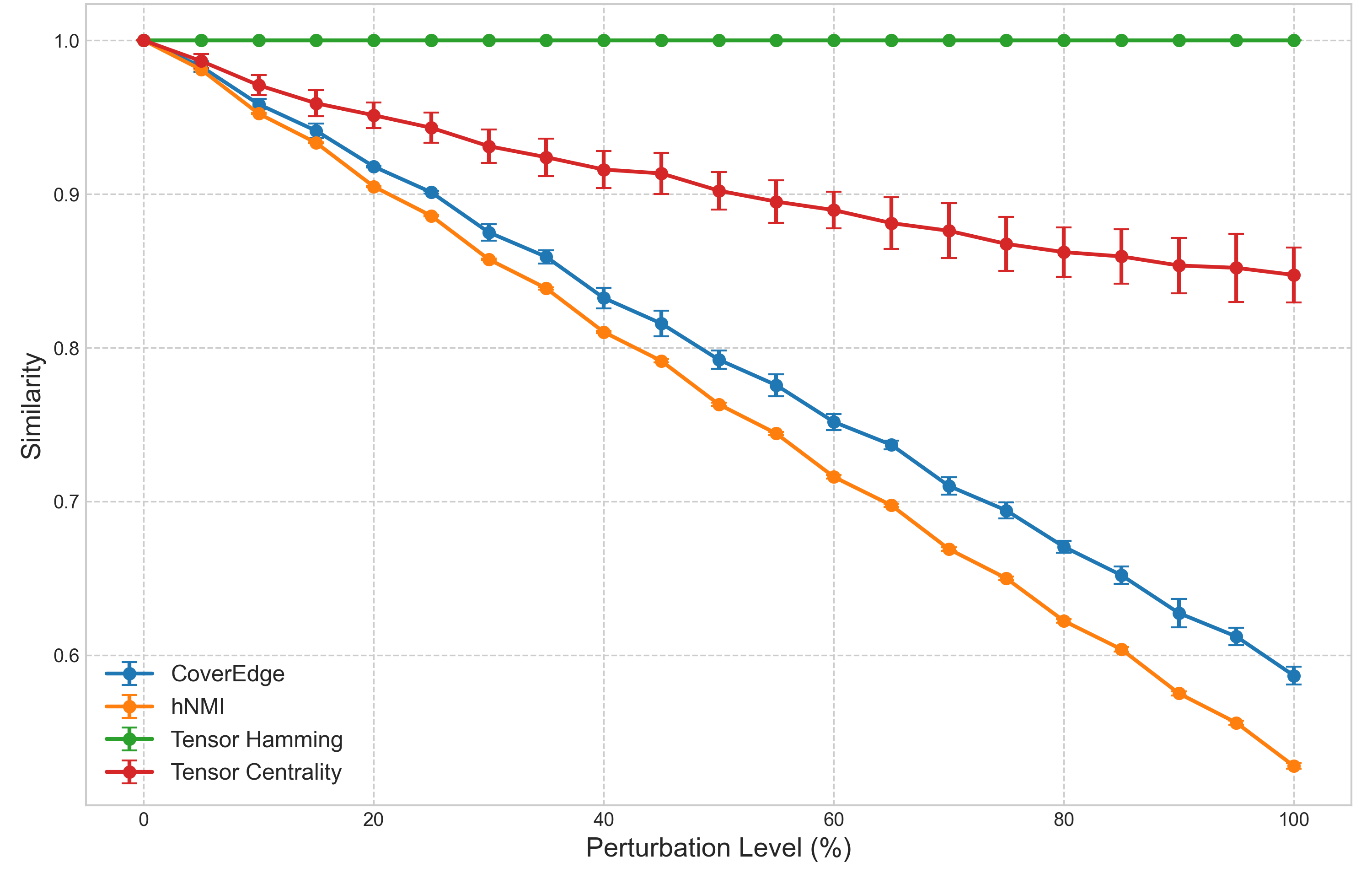}
  \caption{%
The difference in scores from hNMI and CES when perturbations are either made only in the small hypredges of the hypergraph, or only in the large hyperedges.%
  }
  \label{fig:rewire}
\end{figure}

For the second experiment, a fraction $p$ of hyperedges is selected for modification. For each selected hyperedge, a single vertex is randomly removed and replaced with another vertex chosen uniformly from the remaining universe of vertices not already in that edge. This procedure preserves the edge cardinality while introducing local changes to connectivity. Figure~\ref{fig:rewire} shows the results.

Both experiments give us similar results, with hNMI following the perturbation percentage almost perfectly and CES deviating only a little. Tensor-Centrality does capture some of the degradation of the hypergraph, but is still giving very high similarity scores even when 100 percent of the hyperedges are replaced by a random set of vertices. Tensor-Hamming fails completely, likely due to the choice of normalizer.

\begin{figure}[tb]
  \centering 
  \includegraphics[width=\columnwidth]{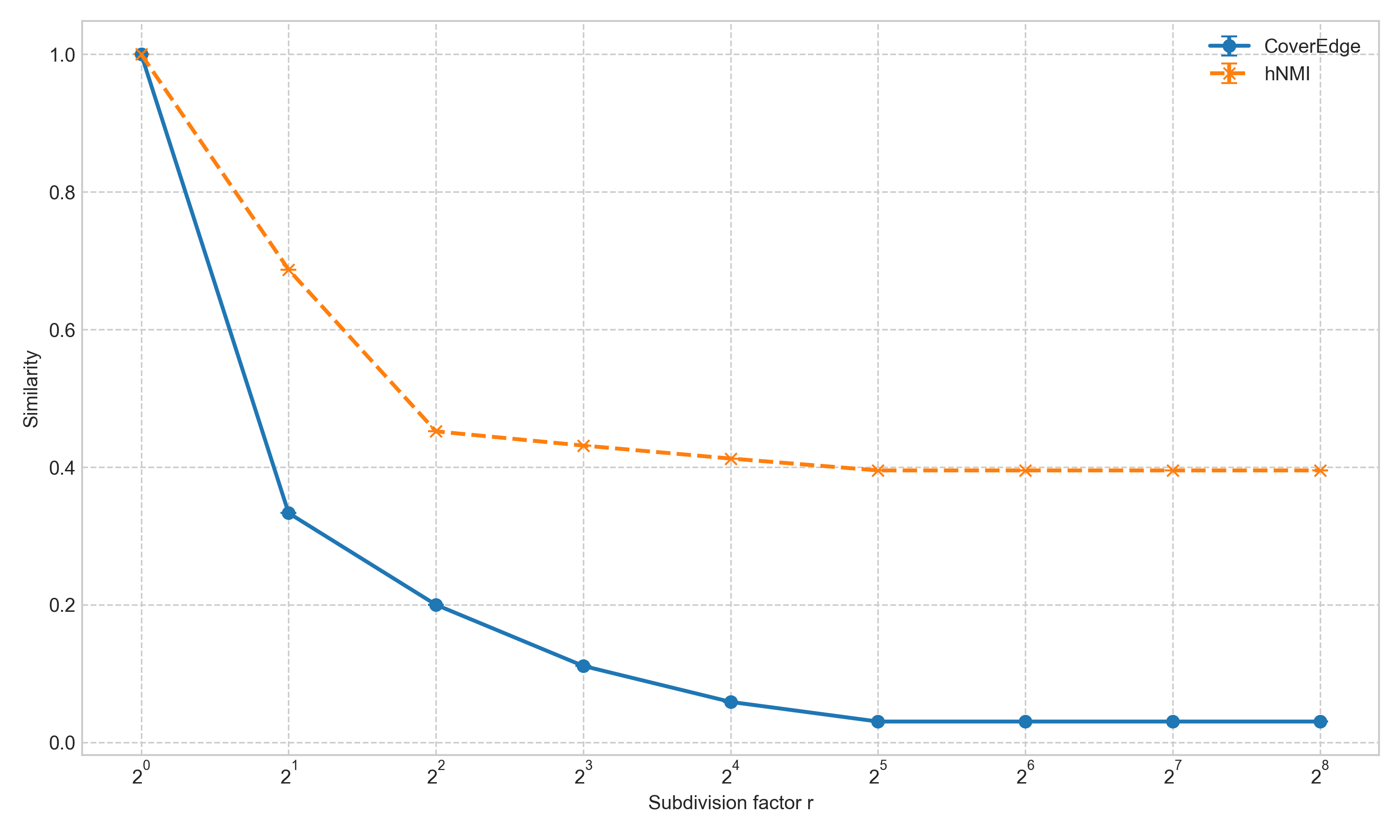}
  \caption{%
The effects of oversegmentation on CES and hNMI.%
  }
  \label{fig:overseg}
\end{figure}

Finally, to model oversegmentation, we split each hyperedge into $r$ smaller parts while keeping the original parent edge. No new vertices are added: the $c$ vertices of each edge are simply shuffled and divided as evenly as possible into $r$ disjoint sub-edges (each of size about $c/r$. If $c<r$, the edge is left unchanged. Increasing $r$ thus increases the number of redundant hyperedges. hNMI does not show problems with oversegmentation, nor does CES. As intended, CES punishes additional redundant hyperedges more strongly than hNMI. Figure~\ref{fig:overseg} shows the results.

We cannot test the Tensor measures for these last two experiments, since the measures for hypergraphs with hyperedges containing 100 vertices well exceed the memory capacity of a 24 GB VRAM GPU. 

\begin{figure}[tb]
  \centering 
  \includegraphics[width=\columnwidth]{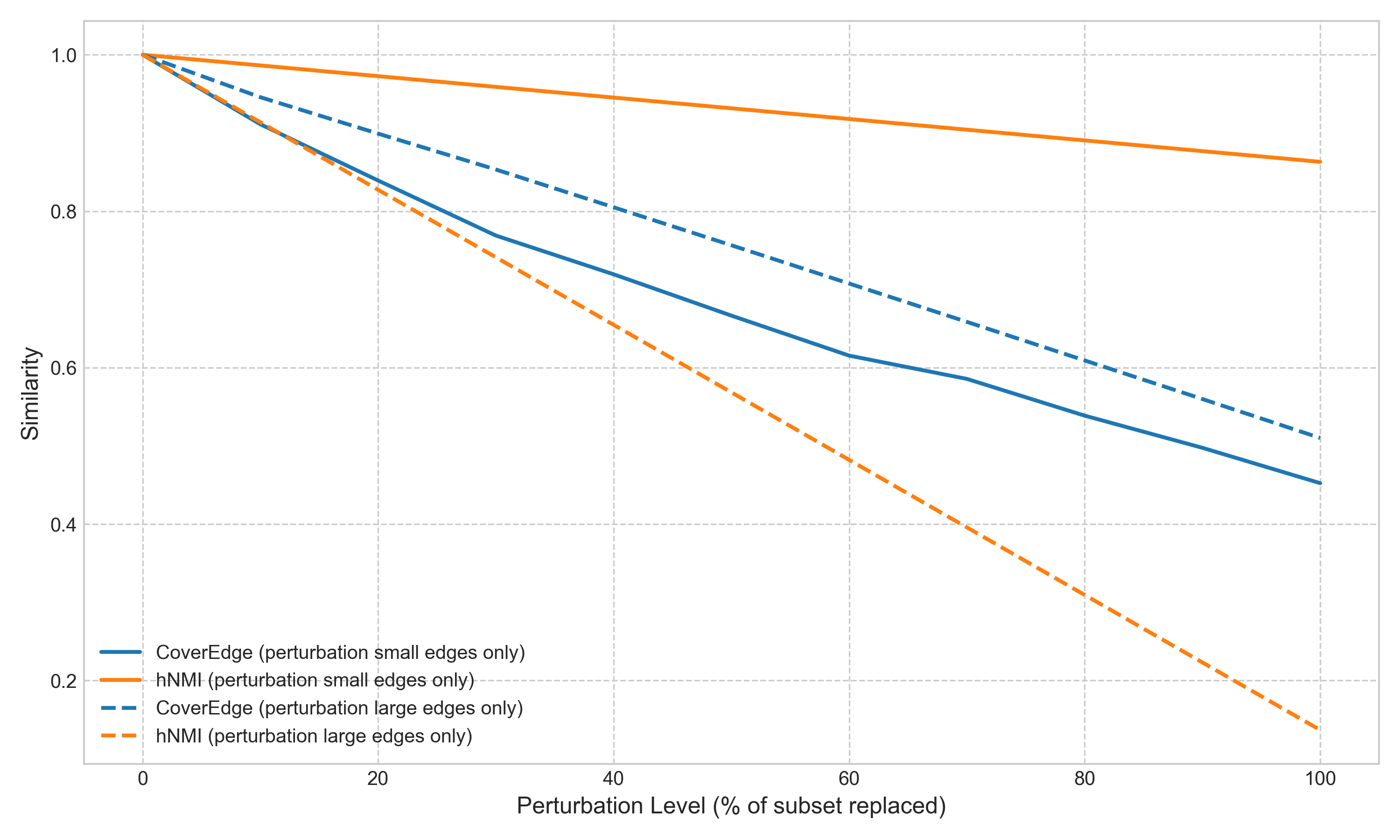}
  \caption{%
The difference in scores from hNMI and CES when perturbations are either made only in the small hypredges of the hypergraph, or only in the large hyperedges.%
  }
  \label{fig:imbalanced}
\end{figure}

\section{Results for distinguishing hypergraph constructors}

Here, we assess the performance of CES and hNMI on distinguishing synthetic hypergraphs. We take the same approach as in \cite{Surana2023HypergraphSimilarityMeasures}, where we construct several hypergraphs with the three different synthetic hypergraph constructors: Erd\H{o}s--R\'enyi (ER), Barab\'asi--Albert (BA), and Watts-Strogatz (WS). A good similarity measure should be able to distinguish a hypergraph generated by one constructor from another. In our experiment, we will compare hNMI with CES.

To quantify the performance of the similarity measures, we use the receiver operating characteristic (ROC) curve. The ROC curve plots the true positive rate against the false positive rate for various threshold values. For a given similarity measure, one selects a threshold > 0 and classifies two hypergraphs as belonging to the same model class if their similarity is above this threshold. Knowing the true classes, true positive and false positive rates can be calculated and varied over thresholds to generate the ROC curve. The area under the curve (AUC) reflects classification performance, with AUC = 1 indicating perfect classification and AUC = 0.5 representing random guessing.

\begin{figure}[tb]
  \centering 
  \includegraphics[width=\columnwidth, alt={Receiver Operator Characteristics curves for three methods.}]{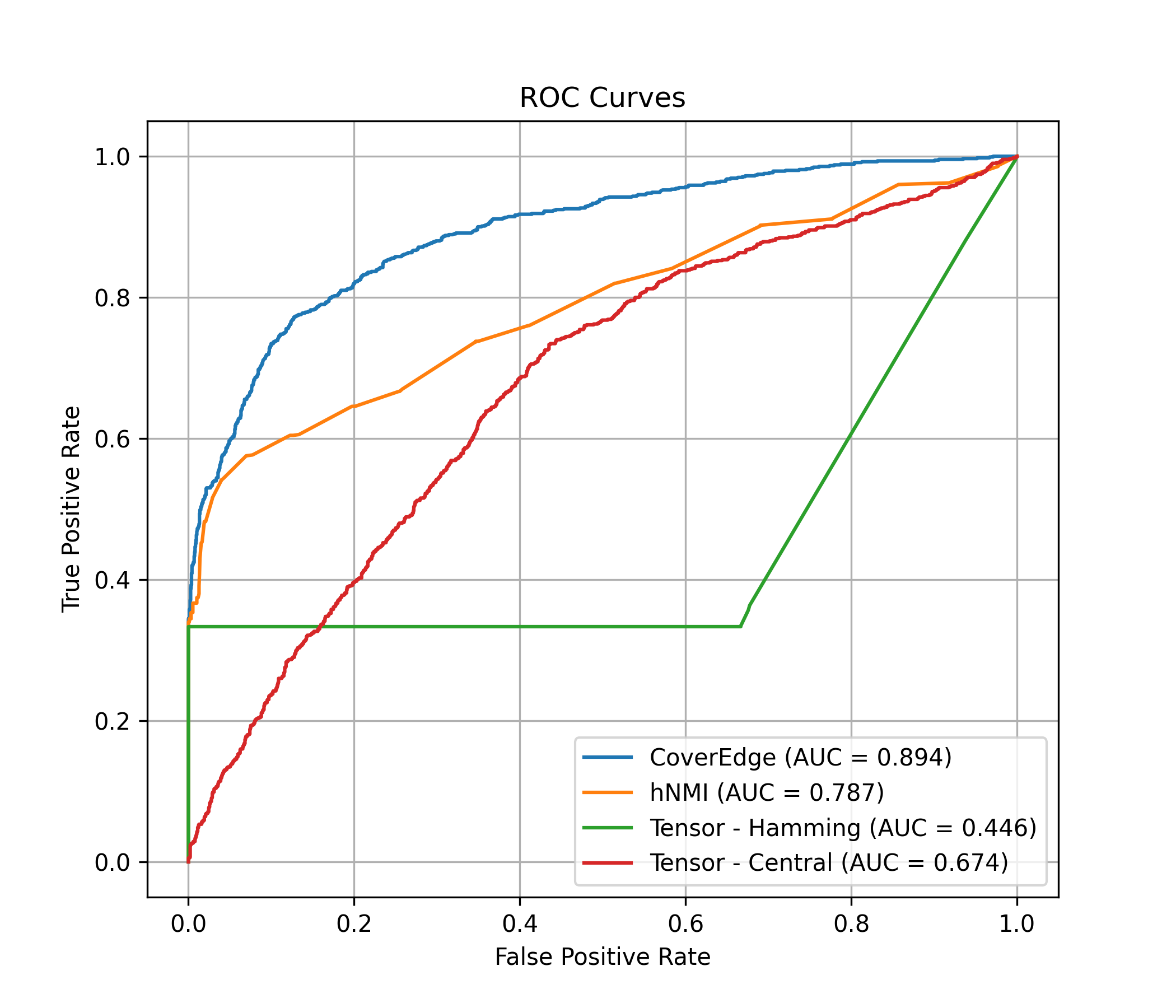}
  \caption{%
ROC and AUC for each similarity measure.%
  }
  \label{fig:roc}
\end{figure}

\begin{figure}[tb]
  \centering 
  \includegraphics[width=\columnwidth, alt={Receiver Operator Characteristics curves for three methods.}]{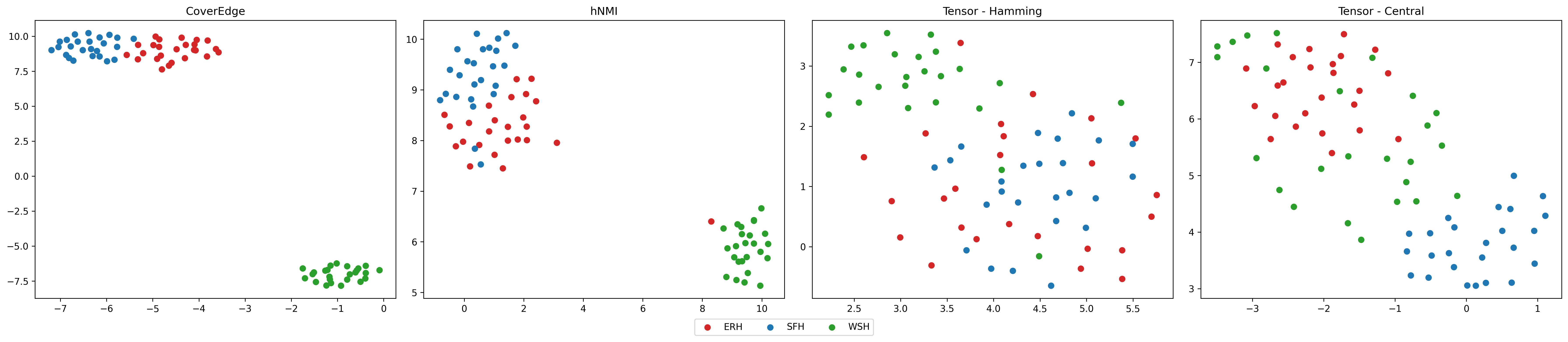}
  \caption{%
UMAP projection using the pairswise similarity results for each measure.%
  }
  \label{fig:umap_synth}
\end{figure}

For evaluation, we generate 25 hypergraphs each for three models, with 40 vertices and 50 hyperedges, yielding a total of 75 graphs. Pairwise similarity values are computed to generate the average ROC curves (Figure~\ref{fig:roc}). Additionally, we apply UMAP to visualize how well each measure is able to separate the construction methods (Figure~\ref{fig:umap_synth}). 

The figures show that CES outperforms the other methods, but that hNMI is also able to separate the different constructors quite well. In our experiment, Tensor-Hamming scores worse than in \cite{Surana2023HypergraphSimilarityMeasures}, but Tensor-Centrality does better. We do not know where this discrepancy comes from. 

\section{Results for ground truth based evaluation for all image collections}
\begin{figure*}[ht]
       \includegraphics[width=\textwidth]{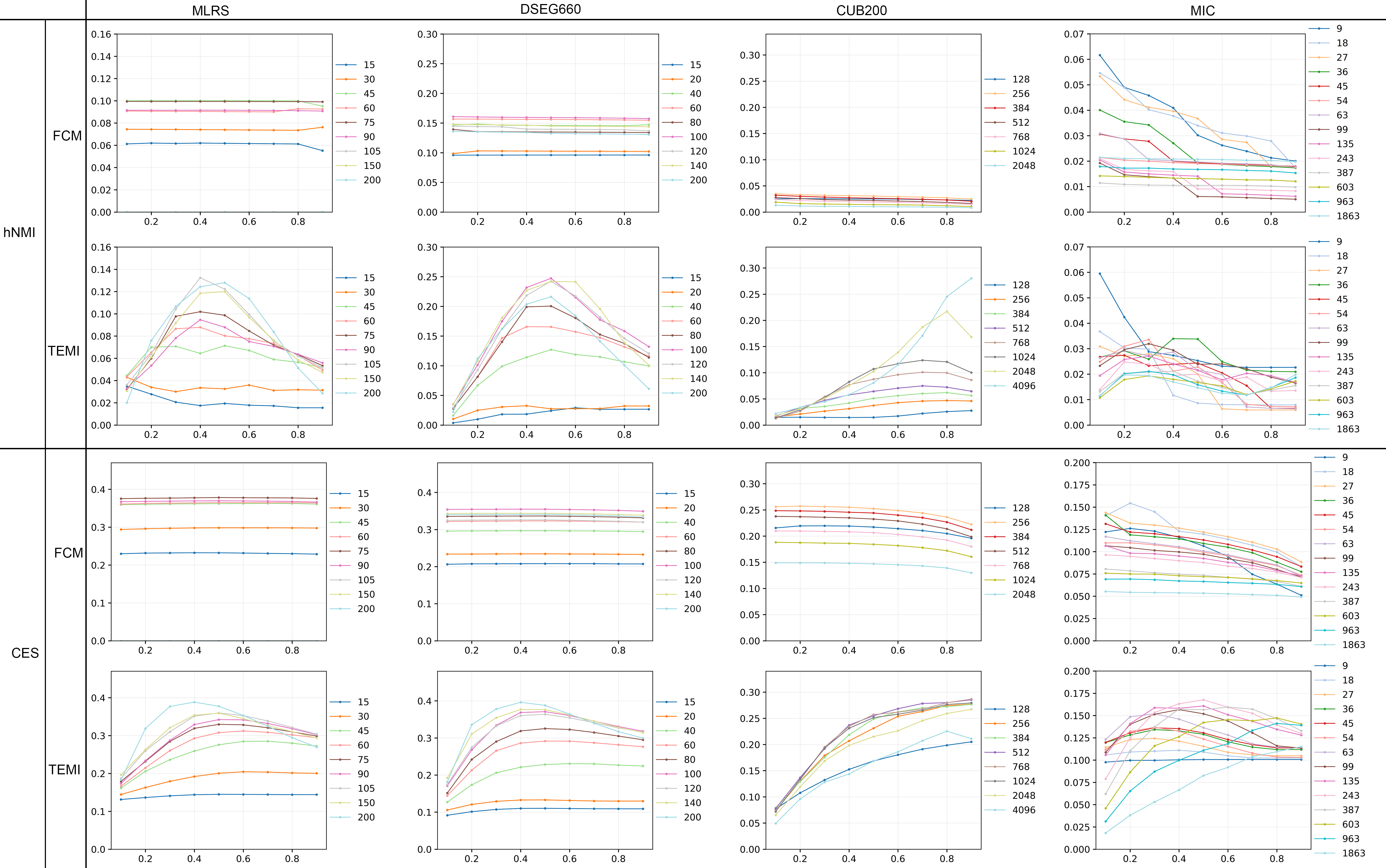}
          \caption{Results for various combinations of $t$ and $k$ for MLRS, DSEG660, CUB200 and MIC, using either FCM or TEMI to construct the hypergraph and hNMI or CES to evaluate the constructed hypergraph against the ground truth hypergraph.}
     \label{fig:allresults2}
   \end{figure*}

In Figure~\ref{fig:allresults2} we show the results for the constructed hypergraphs by FCM and TEMI on each image collection, compared to their ground truth. This includes the results for CUB200 and MIC, which are also shown in the main paper. 

The figure shows that TEMI generally outperforms FCM. hNMI and CES show fairly similar results for the best combination of $k$ and $t$, except for CUB200, where hNMI continues to increase with larger values of $k$.

\section{User evaluation}
This section contains the task list for each image collection used during the user evaluation. For each task, a detailed description of the participants’ various approaches is provided. In the MIC and MAIC evaluations, three domain experts from the aviation and marine sectors participated, respectively. For PAIC, two experts from the construction and fire departments participated. All participants were affiliated with the Dutch Safety Board. The datasets were obtained as part of official investigations conducted by the Dutch Safety Board and are therefore classified.

\subsection{Tasks and approaches for the MH17 image collection (MIC)}
This image collection consists of 12,000+ images collected by the Dutch Safety Board during the investigation of the MH17 crash in 2014.

\begin{itemize}
\item \textbf{Task 1 - 10 minutes} 
Locate and explore the cockpit wreckage site. Describe what you see as accurately and in as much detail as possible. Describe the site (location), not the cockpit itself. 

P5 used the overview to find hyperedges that could possibly have images related to the cockpit. They knew this particular airplane had a specific color used in the cockpit interior that was not used anywhere else in the airplane. This helped to identify a hyperedge that did indeed contain images of the cockpit. They queried an image that showed a decent amount of the surrounding area to find more such images. Combining the visual information in those images, they were able to give an accurate description of the location (a location within a sunflower field, near an old, slightly ruined looking building, with a row of planted trees nearby, a hill in the background and telephone or electricity poles further into the field.

P6 used the text query to find images of the cockpit. While it did yield images of intact cockpits, from other airplanes, this was not was they were looking for. They switched to the Overview, but were unable to detect any cockpit images from there. They tried the text query again, and inspected some of the images more closely. They noted that to start, they wanted to find an image from which they could then extrapolate their search, and that the overview and the text query seemed the most logical way to find such an image. They switched again to the Overview, and found a hyperedge that seemed to show fuselage that also contained the window sills of the cockpit window, a characateristic shape for the trained eye. Since the task did not give a definition of 'site', the user assumed the location of this part of the airplane was the cockpit site. Since the image they found did not show much of the environment, they used query by selection to find additional images. Some showed more of the environment. They added the images of the object found so far to a hyperedge, and proceeded using the Spatial view to find even more images that were linked to this newly created hyperedge. Combining the information gathered from the set of images, they gave a clear and accurate description of the location of the object. 

P8 also used the text query to search for cockpit. They then used the intersection list adjacent to the Hyperedge grid to see a lot of their query results were in edge 62. They explore this edge, create a cockpit hyperedge and adds several of the found images. They use the intersection list again to find another edge with relevant images and proceeds to add more images of the cockpit to the custom hyperedge. They then use query by selection to further enhance their hyperedge. With those results, they found the same location as P5, although they were not able to give as accurate a description, as they were not able to find images that clearly depicted the environment.

\item \textbf{Task 2 - 10 minutes}
Find all, or as many as possible, images that show different parts of the fuselage of the aircraft. Collect them into a hyperedge. 

To find the different fuselage images, P5 first used to overview to find any obvious hyperedges. Opening these and adding images to a new "fuselage" hyperedge. He then used the Spatial hyperedge view to find hyperedges linked to this fuselage hyperedge, and used the lasso tool to select images nearby linked images, to find additional pieces of fuselage. He noted that while there are some ways to see which images have been identified already, the Spatial view does not show that. In total, P5 found 42 different pieces of fuselage.

P6 uses text query to search for "fuselage". This results in several useful images. They add it to their newly created fuselage hyperedge. They then use the query hyperedge to find additional images. From there, they move on to the spatial view. They repeat these steps several times, and within 10 minutes they found 705 fuselage images, of which 41 were unique parts.

P8 starts with the already opened hyperedge, since it showed some fuselage. They make the new edge and adds the images they found there. They use query by selection to find more and add more. They found one image of fuselage that looked different from the rest, queried that one specifically to find more different parts of the fuselage. They continued this process with finding the most different looking correct results, and querying those. Like P6, P8 also added many images of the same fuselage, and found 616 images of fuselage in total, of which 28 were unique parts.

\item \textbf{Task 3 - 10 minutes}
Identify any images that appear to be completely unrelated to MH17. 

P5 said it just makes sense to start with the overview, and used it to find obviously unrelated images. they then used the text query to search for some terms that would likely yield results unrelated to MH17 (if they were in MIC), such as "tropical" and "logo". In 5 minutes, P5 was able to find 654 images unrelated to the MH17 crash.

P6 started with the Overview as well, and found several hyperedges that contain clearly irrelevant images. Just using the Overview, after about 10 minutes, they have found 875 images they deemed irrelevant. 

P8 remembered that the very first hyperedge in the Hyperedge list had some irrelevant images, and used those to query by selection to find more. Any time they found an irrelevant item that looked different from what they already had, they used the query by selection again. When this was exhausted, they opened the next hyperedge and repeated the process. In total, they found 1955 irrelevant images.

\item \textbf{Task 4 - until found}
Investigate whether there is evidence for the presence of shrapnel-like perforations in the cockpit and its fuselage. 

P5 used the overview to identify initial images showing impact damage. They then switched, because they remembered already adding some images like that to the fuselage hyperedge from task 2. They query some of the images and identified a cockpit window fuselage part with impact damage. It took P5 about 2 minutes.

Since P6 already found images of the cockpit containing shrapnel damage, they had solved the fourth task by default.

P8 used the earlier created fuselage hyperedge, because they remembered there were objects with perforations. They made a shrapnel hyperedge and added some images of fuselage containing some damage patterns that do not seem related to crash impact or fire. They perform a hyperedge query and finds more images deemed relevant. After inspecting the resulting hyperedge, they recognizes a part of the fuselage as part of the cockpit. 

\end{itemize}

We found that each participant took a different but successful approach to the tasks. Participants frequently switched between different views, as well as between exploratory and targeted search strategies. The tasks were designed such that the completion of one could inform the next, reflecting the iterative nature of real-world investigations. However, this structure rendered some tasks overly trivial for certain participants.

\subsection{Tasks and approaches for the Marine accident image collection (MAIC)}
This image collection consists of 42,960 images related to various incidents investigated by the Dutch Safety Board over the past 10 years. 

\begin{itemize}
\item \textbf{Task 1 - 10 minutes} 
Make a hyperedge named Ship types. Find and collect one image for each different ship type in the image collection. 

For the first task, P1 and P2 took a similar approach, starting with the Overview to find hyperedges displaying different ship types. In both cases there were enough relevant results that they did not need to use other methods before the allotted time was over. They found 22 and 24 respectively. P3 used a different approach, using the text query to find different ship types. This too worked well, and shows that domain expertise helps in this approach, being familiar with all the different types of ships. He found 17 ship types.

\item \textbf{Task 2 - 10 minutes} 
Locate and collect as many images as you can that show heavy damage to the hull of a ship. 

To find as many images of hull damage as possible, P1 started with the overview, and found one hyperedge where the core images clearly depicted damage to the hull of a ship. Opening the hyperedge on the Hyperedge image grid revealed more images of the same ship with the damaged hull, including both images showing the damaged hull and other images. After adding the relevant images to a newly created hyperedge they named "Hull damage", they went on to query the hyperedge. This revealed several more relevant images, that were also added to the new hyperedge. However, P1 was not quite satisfied with the accuracy of the retrieved images from the query, as it mostly showed intact ships from a similar angle as most of the images in the queried hyperedge. Opening one of the images with clear hull damage, they selected the part of the image that was severely damaged and queried it. This resulted in quite a few more images of different ships with large gashes in their hull. Repeating this process a couple of times for the images that showed somewhat different damage patterns, the hit rate remained fairly high until the time for the task was over. In the 5 minutes allotted, P1 found 89 relevant images. 
P2 and P3 took a fairly similar approach to each other. For them, the text query was very intuitive, and seemed to work well for most tasks. They used both Image query and ROI query to find additional images. P2 found 45 images, and P3 found 15 images, however, it should be noted that P3 decided to only add unique ships to the custom hyperedge.

\item \textbf{Task 3 - until found} 
Find images showing an image taken from the bridge of a fishing vessel. 

P1 found an example of an image taken from the bridge of a fishing vessel within 3 minutes. First, they used the overview to find a hyperedge of images taken from the bridge of a ship. There seemd to be only one hyperedge clearly focused on these kinds of images. However, after assessing the images in this hyperedge, they were mostly images of containerships. P1 used the Spatial hyperedge view to see if there were any other hyperedges with a fairly large number of links going towards it. There were a few obvious candidates. Zooming in on one of them revealed a cluster of image nodes within the hyperedge node with quite a few links going towards them. P1 used the lasso tool to select and display the images on the Hyperedge image grid. Since it required a closer inspection to notice the difference between one type of ship's bridge and another, P1 opened the image pop up to display a full sized image, and started to browse through each image. After 15 images, P1 found an image taken from the bridge of a fishing vessel, clearly showing the nets in view through the windows. The hyperedge where the user found the image was fairly broad, containing images with any kind of control panel or other intricate close up electronics, which is why it was not clear from the overview that this might be the right candidate for closer analysis. 

P2 tried a text query, containg bridge and fishing vessel, and it returned somewhat relevant images, either depicting bridges of fishing vessels from outside, or images from inside a ship's bridge (but not a fishing vessel). They then turned to the overview and found a hyperedge containing many images taken from the bridge. After inspecting some and not finding a correct image, they went back to the text query and tried various combinations to no result. Finally, returning to the earlier found hyperedge and going through its images and inspecting them closely, they found a correct image. It took about 18 minutes.

P3 went so far as to copy the whole task question of task 3 into the query, and, to our surprise, the first image of the query results was a correct image. It took P3 less than 10 seconds to find the correct image.

\item \textbf{Task 4 - until found} 
Find an image of a broken mast of a sailing vessel.

P1 used the overview to find images of sailing vessels, then used ROI query of a mast specifically. This lead to correct images. It took him about 5 minutes.

P2 and P3 used the text query again. Due to slightly different queries, P3 found results almost immediately, taking less than one minute of perusing the query results. P3 went on to search more images, since the images found did not directly depict the mast being from or on a ship, but was unable to find an image to verify. P2 required several query attempts, and took about 3 minutes to find a correct image.

\item \textbf{Task 5 - until found} 
Find a ship on a non-Dutch river. 

For task 5, P1 used the overview again and found a hyperedge containing some images on waterways. However, none were clearly rivers or clearly outside of the Netherlands. Using the Spatial hyperedge view, they found a group of images and used the lasso tool. After perusing several of the images, they found one where some cars were parked along a river with a ship on it. The cars had clear non-Dutch license plates. This took about 4 minutes.

P2 used the text query again to find initial relevant images, but was unable to directly find a correct image. Using Image query, they found additional images. Repeating this a couple of times, they eventually found an image with a ship on a river with a snowy riverbank. Noting this is not a common sight in the Netherlands, he (correctly) assumed a correct image was found. They took around 10 minutes.

P3 used a similar approach, and in fact found the same image as P2 within a couple of minutes. However, not satisfied with being unable to verify that it was indeed not in the Netherlands, they continued searching. Through various queries, they found a different correct image with a verifiable non-Dutch location. It took about 12 minutes.

\end{itemize}

The combination of the MAIC and tasks set out by us, show that the Spatial view can be useful, but was not necessary. In fact, most tasks could be solved by standard retrieval methods. This was a combination of underestimating the power of the OpenCLIP features by the authors, and the query ingenuity by the participants. Although both P2 and P3 did stumble upon some limitations of that combination as well, thus respectively requiring more time for task 3 and 5 than P1, who mostly used the Overview and Spatial view.

\subsection{Tasks and approaches for the Parking garage accident image collection (PAIC)}
This image collection was created during the investigation of the collapse of the ramps of a parking garage in the Netherlands. The images consist mainly of photos made by drones and photos made by investigators and other authorities involved at the parking garage. While this collection is not as large as the others (4,417 images), it is challenging due to each floor of the parking garage looking very similar.
\begin{itemize}
\item \textbf{Task 1 - 10 minutes} 
Remove all images you deem irrelevant to the collapse of the parking garage. 

To remove any irrelevant images, P4 used the overview to see if there were any hyperedges or images that immediately stood out. This was the case, and the participant opened these hyperedge, assessed the images, queried them to retrieve all similar images, and proceeded to remove them from all hyperedges they were part of. P4 removed 94 images.

P7 Started task one by using the Spatial view, with the assumption that images towards the edges of the hyperedge nodes may be irrelevant outliers. While they found that compared to the other images in the nodes, they were indeed outliers, it did not mean they were irrelevant. P7 then switched to the Overview and found several images for the task. After about 10 minutes, they removed 150 irrelevant images.  

\item \textbf{Task 2 - until found} 
Find images depicting the connection between two segments of the ramp beams.  
P4 used overview to find images of the beams. Adding them to a new "beams" hyperedge and then querying the hyperedge resulted in more images. One seemed to depict the connection between the beams, but from a large distance. They used ROI query which resulted in closer up images, confirming that the correct images were found. It took about 1 minute.

P7 started with a text query for "beam", but with unsatisfactory results. They created a hyperedge for their task. P7 switched to the Overview, and started opening some of the hyperedges that looked relevant. They added images with beams to their hyperedge, with the idea to later inspect them more closely for the specific request of the task. They found quite a lot of relevant photos in a few minutes, and then proceeded with the closer inspection. The first photo they checked already contained the requested object. P7 noted that they could have started with inspecting right away and found it much sooner. It took 6 minutes.

\item \textbf{Task 3 - until found} 
Find a close-up of a side view of the third floor, the fourth floor and the fifth floor.  

P7 started with the Overview to find some images that showed the parking garage from a distance. They wanted to use these images to then match close up images with the correct floor. By using ROI query, zooming in on the relevant floors, they found many close up images. However, to their disappointment, they did not exactly match their selected ROI. The model was not able to distinguish well between one floor with rubble and the next. By manually going through some of the query results, the approach still proved successful, and within a 7 minutes they found several relevant images for each floor.
P4 took exactly the same approach, and took about 4 minutes.

\item \textbf{Task 4 - until found} 
Find two distinct hyperedges that show overlap due to a particular object. 

P4 used the Hypergraph matrix by hovering over edges with a low number of intersecting items. Their idea was that these edges would be distinct from  each other, but have for one image something in common. They found two edges, where one hyperedge contained mostly images taken from a distance, whereas the other edge contained close ups of a specific object type, one of which was also present in the distant images. It took about one minute.

P7 browsed through the hyperedge using the Hyperedge list and found close up images of a rubble, with in the background a grid fence. This same type of fence also showed up in another hyperedge, containing images of the top floor. It took P7 2 minutes.
\end{itemize}

\end{document}